\documentclass[10pt,letterpaper,english]{amsart}

\usepackage{array}
\usepackage{subfigure}
\usepackage{graphicx}
\usepackage{amssymb}

\makeatletter

\providecommand{\tabularnewline}{\\}

 \theoremstyle{plain}    
 \newtheorem{thm}{Theorem}[section]
 \numberwithin{equation}{section} 
 \numberwithin{figure}{section} 
 \theoremstyle{plain}
 \theoremstyle{definition}
 \newtheorem{defn}[thm]{Definition}
 \theoremstyle{remark}    
 \newtheorem{claim}[thm]{Claim}
 \theoremstyle{plain}    
 \newtheorem{conjecture}[thm]{Conjecture} 
 \theoremstyle{plain}    
 \newtheorem{lem}[thm]{Lemma} 
 \theoremstyle{remark}    
 \newtheorem*{acknowledgement*}{Acknowledgement} 

\DeclareMathOperator{\diag}{diag}
\DeclareMathOperator{\Ai}{Ai}
\DeclareMathOperator{\tr}{tr}
\DeclareMathOperator{\betapdf}{beta}
\DeclareMathOperator{\Var}{Var}

\makeatother
\begin{document}

\title{From Random Matrices to Stochastic Operators}

\author{Alan Edelman and Brian D. Sutton}

\keywords{random matrices, random eigenvalues, stochastic differential operators}

\thanks{Submitted for publication 19 July 2006. Revised 6 September 2006.}

\address{Alan Edelman\\
Department of Mathematics \\
Massachusetts Institute of Technology\\
Cambridge, MA\ \ 02139}

\address{Brian D. Sutton\\
Department of Mathematics\\
Randolph-Macon College\\
Ashland, VA\ \ 23005}

\begin{abstract}
We propose that classical random matrix models are properly viewed
as finite difference schemes for stochastic differential operators.
Three particular stochastic operators commonly arise, each associated
with a familiar class of local eigenvalue behavior. The \emph{stochastic
Airy operator} displays soft edge behavior, associated with the Airy
kernel. The \emph{stochastic Bessel operator} displays hard edge behavior,
associated with the Bessel kernel. The article concludes with suggestions
for a \emph{stochastic sine operator}, which would display bulk behavior,
associated with the sine kernel.
\end{abstract}
\maketitle

\section{Introduction}

Through a number of carefully chosen, eigenvalue-preserving transformations,
we show that the most commonly studied random matrix distributions
can be viewed as finite difference schemes for stochastic differential
operators. Three operators commonly arise---the stochastic Airy, Bessel,
and sine operators---and these operators are associated with three
familiar classes of local eigenvalue behavior---soft edge, hard edge,
and bulk.

For an example, consider the Hermite, or Gaussian, family of random
matrices. Traditionally, a random matrix from this family has been
defined as a dense Hermitian matrix with Gaussian entries, but we
show that such a matrix is equivalent, via similarity, translation,
and scalar multiplication, to a matrix of the form\[
\frac{1}{h^{2}}\Delta+\diag_{-1}(x_{1},\dots,x_{n-1})+\frac{2}{\sqrt{\beta}}\cdot\text{``noise''},\]
in which $\Delta$ is the $n$-by-$n$ second difference matrix, $\diag_{-1}(x_{1},\dots,x_{n-1})$
is an essentially diagonal matrix of grid points, and the remaining
term is a random bidiagonal matrix of {}``pure noise.'' We claim
that this matrix encodes a finite difference scheme for \[
-\frac{d^{2}}{dx^{2}}+x+\frac{2}{\sqrt{\beta}}\cdot\text{``noise''},\]
which is the inspiration for the stochastic Airy operator. (The {}``noise''
term will be made precise later.)

The idea of interpreting the classical ensembles of random matrix
theory as finite difference schemes for stochastic differential operators
was originally presented in July 2003 \cite{siamconference}, and
the theory was developed in \cite{mythesis}. The present article
contains several original contributions, including firm foundations
for the stochastic Airy and Bessel operators.

\bigskip{}

The standard technique for studying local eigenvalue behavior of a
random matrix distribution involves the following steps. (1) Choose
a family of $n$-by-$n$ random matrices, $n=2,3,4,\dots$, (2) Translate
and rescale the $n$th random matrix to focus on a particular region
of the spectrum, and (3) Let $n\rightarrow\infty$. When this procedure
is performed carefully, so that the eigenvalues near zero approach
limiting distributions as $n\rightarrow\infty$, the limiting eigenvalue
behavior often falls into one of three classes: soft edge, hard edge,
or bulk.

\begin{figure}
\begin{center}\subfigure[soft edge]{\includegraphics{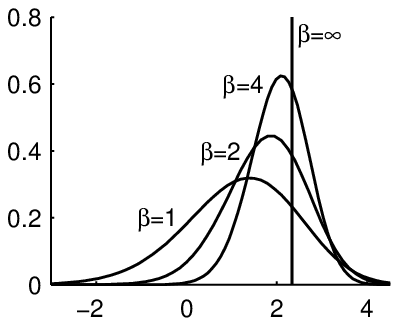}}\subfigure[hard edge ($a=0$)]{\includegraphics{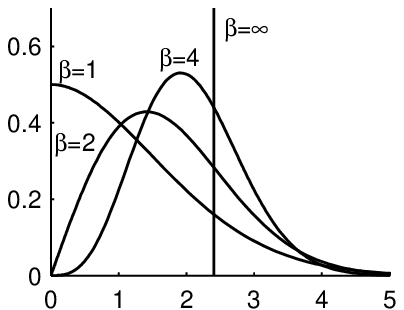}}\subfigure[bulk]{\includegraphics{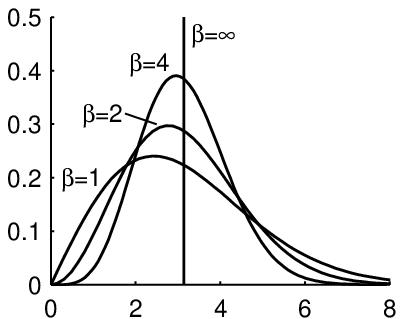}}\end{center}

\caption{\label{fig:softhardbulk}Soft edge, hard edge, and bulk distributions,
associated with the Airy, Bessel, and sine kernels, respectively.}
\end{figure}

The largest eigenvalues of many random matrix distributions, notably
the Hermite (i.e., Gaussian) and Laguerre (i.e., Wishart) ensembles,
display soft edge behavior. The limiting marginal density, as the
size of the matrix approaches infinity, of a single eigenvalue at
the soft edge is associated with the Airy kernel. Tracy and Widom
derived formulas for these density functions in the cases $\beta=1,2,4$,
relating them to solutions of the Painlev\'{e} II differential equation.
See Figure \ref{fig:softhardbulk}(a). Relevent references include
\cite{MR1236195,MR1737991,MR1863961,tracywidomletter,MR1257246,MR1385083,MR1844228}. 

The smallest eigenvalues of some random matrix distributions, notably
the Laguerre and Jacobi ensembles, display hard edge behavior. The
limiting marginal density of a single eigenvalue at the hard edge
is associated with the Bessel kernel. Formulas exist for these density
functions as well, expressible in terms of solutions to Painlev\'{e}
equations. See Figure \ref{fig:softhardbulk}(b). Relevant references
include \cite{MR1236195,smallesteigenvalueinlaguerre,MR1912278,MR1266485}. 

The eigenvalues in the middle of the spectra of many random matrix
distributions display bulk behavior. In this case, the spacing between
consecutive eigenvalues is interesting. The spacing distributions
are associated with the sine kernel, and formulas for the density
functions, due to Jimbo, Miwa, M\^{o}ri, Sato, Tracy, and Widom,
are related to the Painlev\'{e} V differential equation. See Figure
\ref{fig:softhardbulk}(c). Relevant references include \cite{MR0573370,MR1912278,MR1083764,MR1253763}.

This article contends that the most natural setting for soft edge
behavior is in the eigenvalues of the \emph{stochastic Airy operator}\begin{equation}
-\frac{d^{2}}{dx^{2}}+x+\frac{2}{\sqrt{\beta}}\cdot\text{``noise''},\label{eq:stochasticairywithquotes}\end{equation}
and the most natural setting for hard edge behavior is in the singular
values of the \emph{stochastic Bessel operator}\begin{equation}
-2\sqrt{x}\frac{d}{dx}+\frac{a}{\sqrt{x}}+\frac{2}{\sqrt{\beta}}\cdot\text{``noise''}.\label{eq:stochasticbesselwithquotes}\end{equation}
A suggestion for a \emph{stochastic sine operator}, along the lines
of (\ref{eq:stochasticairywithquotes}--\ref{eq:stochasticbesselwithquotes}),
is presented at the end of the article. The correct interpretations
of the {}``noise'' terms in (\ref{eq:stochasticairywithquotes})
and (\ref{eq:stochasticbesselwithquotes}) will be specified later
in the article, as will boundary conditions; see Definitions \ref{def:stochasticairy}
and \ref{def:stochasticbessel}. The parameter $\beta$ has its usual
meaning from random matrix theory, but now the cases $\beta=1,2,4$
do not seem special.

\begin{figure}
\begin{center}\subfigure[$\beta=4$]{\includegraphics{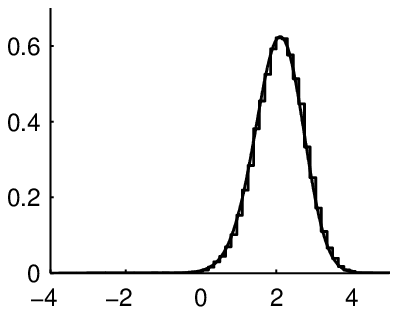}}\subfigure[$\beta=2$]{\includegraphics{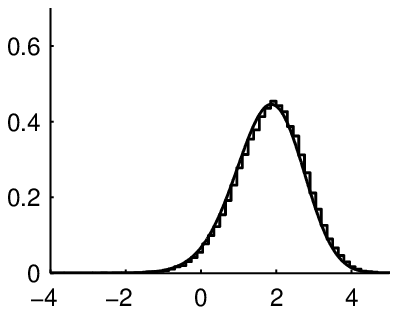}}\subfigure[$\beta=1$]{\includegraphics{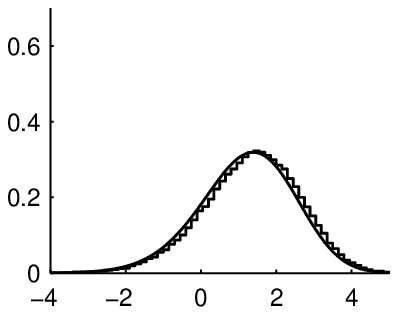}}\end{center}

\caption{\label{fig:airyrayleighritz}Least eigenvalue of the stochastic Airy
operator. In each plot, the smooth curve is a soft edge (Tracy-Widom)
density, and the jagged curve is a histogram of the least eigenvalue
from $10^{5}$ random samples of the stochastic Airy operator. The
small positive bias in each histogram results from the use of a Rayleigh-Ritz
procedure, which overestimates eigenvalues.}
\end{figure}

\begin{figure}
\begin{center}\subfigure[$\beta=4,\;a=0$]{\includegraphics{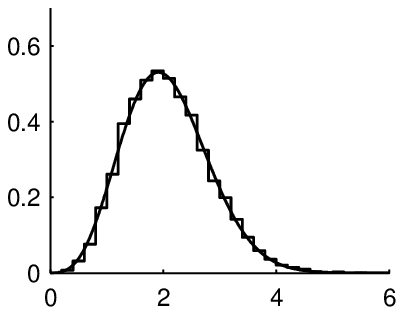}}\subfigure[$\beta=2,\;a=0$]{\includegraphics{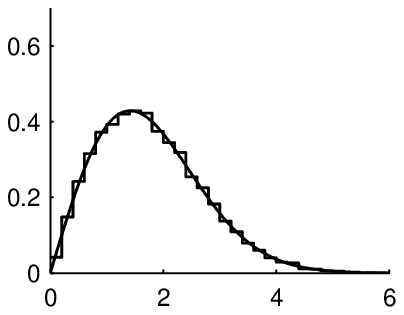}}\subfigure[$\beta=1,\;a=0$]{\includegraphics{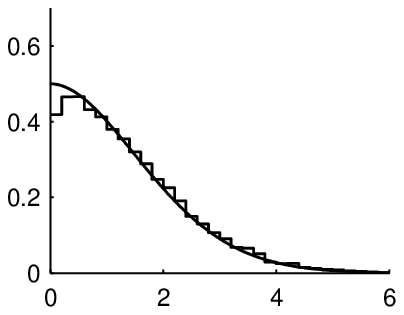}}\end{center}

\caption{\label{fig:besselrayleighritz}Least singular value of the stochastic
Bessel operator. In each plot, the smooth curve is a hard edge density,
and the jagged curve is a histogram of the least singular value from
$10^{4}$ random samples of the stochastic Bessel operator. The small
positive bias in each histogram results from the use of a Rayleigh-Ritz
procedure, which overestimates singular values.}
\end{figure}

Numerical evidence is presented in Figures \ref{fig:airyrayleighritz}
and \ref{fig:besselrayleighritz}. The first compares histograms of
stochastic Airy eigenvalues to the soft edge distributions of Figure
\ref{fig:softhardbulk}(a), and the second compares histograms of
stochastic Bessel singular values to the hard edge distributions of
Figure \ref{fig:softhardbulk}(b). The computations were based on
the Rayleigh-Ritz method. They are explained in further detail in
Sections \ref{section:airyrayleighritz} and \ref{section:besselrayleighritz}.

The stochastic Airy, Bessel, and sine operators were discovered by
interpreting the classical ensembles of random matrix theory as finite
difference schemes. We argue that

\begin{enumerate}
\item When scaled at the soft edge, the Hermite and Laguerre matrix models
encode finite difference schemes for the stochastic Airy operator.
\item When scaled at the hard edge, the Laguerre and Jacobi matrix models
encode finite difference schemes for the stochastic Bessel operator.
\end{enumerate}
See Section \ref{sec:discretizationclaims} for an overview. Exactly
what is meant by {}``scaling'' will be developed later in the article.
Typically, scaling involves subtracting a multiple of an identity
matrix and multiplying by a scalar to focus on a particular region
of the spectrum, along with a few tricks to decompose the matrix into
a random part and a nonrandom part. The structured matrix models introduced
by Dumitriu and Edelman \cite{MR1936554} and further developed by
Killip and Nenciu \cite{killipnenciu} and Edelman and Sutton \cite{jacobimodel}
play vital roles.

The original contributions of this article include the following.

\begin{figure}
\begin{center}\begin{tabular}{>{\centering}m{0.65in}|>{\centering}p{1.15in}|>{\centering}p{1.15in}|>{\centering}p{1.15in}}
eigenvector of $H_{\text{soft}}^{\beta}$&
$\log|v|$&
$\nabla\log|v|$&
$\nabla^{2}\log|v|$\tabularnewline
\hline
first&
\includegraphics{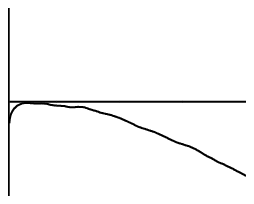}

once differentiable&
\includegraphics{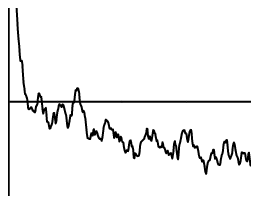}

continuous&
\includegraphics{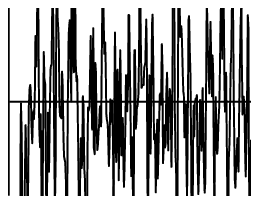}\tabularnewline
\hline
second&
\includegraphics{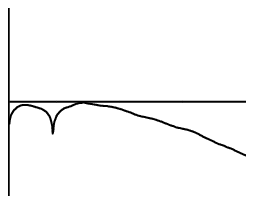}

once differentiable&
\includegraphics{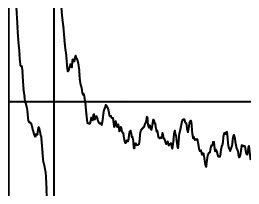}

continuous&
\includegraphics{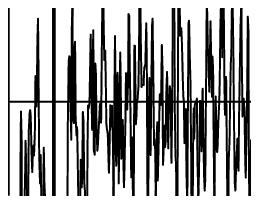}\tabularnewline
\hline
${\displaystyle \frac{\text{second}}{\text{first}}}$&
\includegraphics{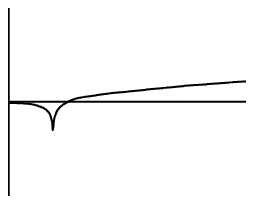}

twice differentiable&
\includegraphics{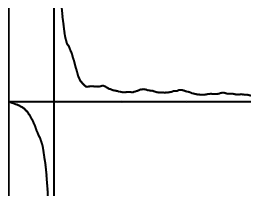}

once differentiable&
\includegraphics{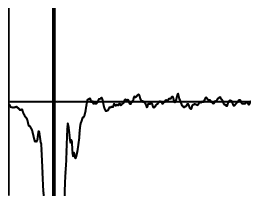}

continuous\tabularnewline
\end{tabular}\end{center}

\caption{\label{fig:predictionforhermite}A prediction of the stochastic operator
approach. The entrywise ratio of two eigenvectors of the rescaled
Hermite matrix model $H_{\text{soft}}^{\beta}$ is {}``smoother''
than either individual eigenvector. The plots are generated from a
single random sample of $H_{\text{soft}}^{\beta}$ with $\beta=2$
and $n=10^{5}$. A log scale is used for visual appeal. $\nabla$
refers to Matlab's \texttt{gradient} function, and $\nabla^{2}$ indicates
two applications of the \texttt{gradient} function. See Section \ref{sec:smoothness}
for details.}
\end{figure}

\begin{figure}
\begin{tabular}{>{\centering}m{1in}|>{\centering}p{1.15in}|>{\centering}p{1.15in}}
right singular vector of $J_{\text{hard}}^{\beta,a,b}$&
$\log|v|$&
$\nabla\log|v|$\tabularnewline
\hline
first&
\includegraphics{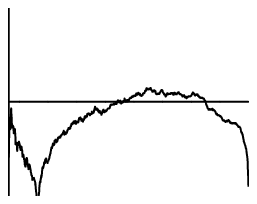}

continuous&
\includegraphics{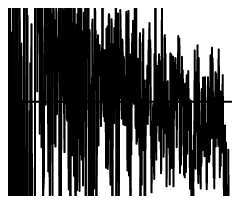}\tabularnewline
\hline
second&
\includegraphics{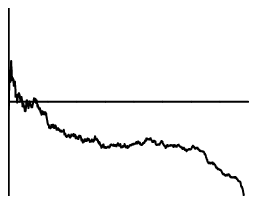}

continuous&
\includegraphics{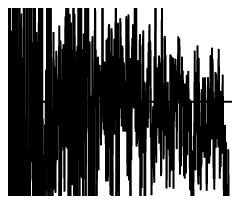}\tabularnewline
\hline
${\displaystyle \frac{\text{second}}{\text{first}}}$&
\includegraphics{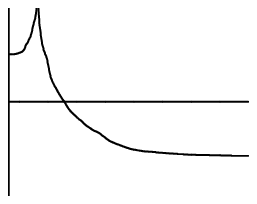}

once differentiable&
\includegraphics{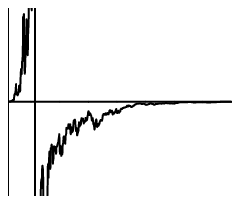}

continuous \tabularnewline
\end{tabular}

\caption{\label{fig:predictionforjacobi}Another prediction of the stochastic
operator approach. The entrywise ratio of two singular vectors of
the rescaled Jacobi matrix model $J_{\text{hard}}^{\beta,a,b}$ is
{}``smoother'' than either individual singular vector. See Section
\ref{sec:smoothness} for details.}
\end{figure}

\begin{itemize}
\item The stochastic Airy and Bessel operators are defined. Care is taken
to ensure that the operators involve ordinary derivatives of well
behaved functions, avoiding any heavy machinery from functional analysis.
\item The smoothness of eigenfunctions and singular functions is investigated.
In the case of the stochastic Airy operator, the $k$th eigenfunction
is of the form $f_{k}\phi$, in which $f_{k}$ is twice differentiable
and $\phi$ is a once differentiable (specifically $C^{3/2-}$) function
defined by an explicit formula. This predicts structure in the eigenvectors
of certain rescaled matrix models, which can be seen numerically in
Figure \ref{fig:predictionforhermite}. Figure \ref{fig:predictionforjacobi}
considers analogous results for the stochastic Bessel operator.
\item The interpretation of random matrix models as finite difference schemes
for stochastic differential operators is developed. This approach
is demonstrated for the soft edge of Hermite, the soft and hard edges
of Laguerre, and the hard edge of Jacobi.
\end{itemize}

Notable work of others includes \cite{MR2172210} and \cite{rideretal}.
Although the stochastic Airy operator is not explictly mentioned in
the large $\beta$ asymptotics of Dumitriu and Edelman \cite{MR2172210},
it appears to play an important role. The stochastic operator approach
has very recently been given a boost by Ram\'{i}rez, Rider, and Vir\'{a}g
\cite{rideretal}, who have proved a conjecture contained in \cite{siamconference,mythesis}
relating the eigenvalues of the stochastic Airy operator to soft edge
behavior. In addition, they have used the stochastic Airy operator
to describe the soft edge distributions in terms of a diffusion process.

The next section reviews necessary background material and introduces
notation. Section \ref{sec:results} provides formal definitions for
the stochastic Airy and Bessel operators and provides an overview
of our results, which are developed in later sections.

\section{\label{sec:background}Background}

Much work in the field of random matrix theory can be divided into
two classes: global eigenvalue behavior and local eigenvalue behavior.

Global eigenvalue behavior refers to the overall density of eigenvalues
along the real line. For example, a commonly studied distribution
on $n$-by-$n$ Hermitian matrices known as the Hermite ensemble typically
has a high density of eigenvalues near zero, but just a scattering
near $\sqrt{2n}$ by comparison. Such a statement does not describe
how the eigenvalues are arranged with respect to each other in either
region, however.

In contrast, local eigenvalue behavior is observed by {}``zooming
in'' on a particular region of the spectrum. The statistic of concern
may be the marginal distribution of a single eigenvalue or the distance
between two consecutive eigenvalues, for example. Local eigenvalue
behavior is determined by two factors---the distribution of the random
matrix and the region of the spectrum under consideration. For example,
the eigenvalues of the Hermite ensemble near zero display very different
behavior from the eigenvalues near the edge of the spectrum, at $\sqrt{2n}$.
Conceivably, the eigenvalues of a different random matrix may display
entirely different behavior.

Interestingly, though, the eigenvalues of many, many random matrix
distributions fall into one of three classes of behavior, locally
speaking. Notably, the eigenvalues of the three classical ensembles
of random matrix theory---Hermite, Laguerre, and Jacobi---fall into
these three classes as the size of the matrix approaches infinity.

In this section, we present background material, covering the three
most commonly studied random matrix distributions and the three classes
of local eigenvalue behavior.

\subsection{\label{sec:matrixmodels}Random matrix models}

There are three classical distributions of random matrix theory\emph{:
Hermite}, \emph{Laguerre}, and \emph{Jacobi}. The distributions are
also called \emph{ensembles} or \emph{matrix models}. They are defined
in this section. Also, joint distributions for Hermite eigenvalues,
Laguerre singular values, and Jacobi CS values are provided. We use
the word \emph{spectrum} to refer to all eigenvalues or singular values
or CS values, depending on context. Also, note that the language earlier
in the article was loose, referring to eigenvalues when it would have
been more appropriate to say {}``eigenvalues or singular values or
CS values.''

\subsubsection{Hermite}

The Hermite ensembles also go by the name of the Gaussian ensembles.
Traditionally, three flavors have been studied, one for real symmetric
matrices, one for complex Hermitian matrices, and one for quaternion
self-dual matrices. In all three cases, the density function is\begin{equation}
\text{const}\times\exp\left(-\frac{\beta}{2}\tr A^{2}\right)dA,\label{eq:gaussianensembledensity}\end{equation}
in which $\beta=1$ for the real symmetric case, $\beta=2$ for the
complex Hermitian case, and $\beta=4$ for the quaternion self-dual
case. The entries in the upper triangular part of such a matrix are
independent Gaussians, although the diagonal and off-diagonal entries
have different variances.

The eigenvalues of the Hermite ensembles have joint density \begin{equation}
\text{const}\times e^{-\frac{\beta}{2}\sum_{i=1}^{n}\lambda_{i}^{2}}\prod_{1\leq i<j\leq n}|\lambda_{i}-\lambda_{j}|^{\beta}\prod_{i=1}^{n}d\lambda_{i}.\label{eq:hermitedensity}\end{equation}

Dumitriu and Edelman extended the Hermite ensembles to all $\beta>0$
\cite{MR1936554}. Below, $X\sim Y$ indicates that $X$ and $Y$
have the same distribution.

\begin{defn}
The $n$-by-$n$ \emph{$\beta$-Hermite matrix model} is the random
real symmetric matrix\[
H^{\beta}\sim\frac{1}{\sqrt{2\beta}}\left[\begin{array}{ccccc}
\sqrt{2}G_{1} & \chi_{(n-1)\beta}\\
\chi_{(n-1)\beta} & \sqrt{2}G_{2} & \chi_{(n-2)\beta}\\
 & \ddots & \ddots & \ddots\\
 &  & \chi_{2\beta} & \sqrt{2}G_{n-1} & \chi_{\beta}\\
 &  &  & \chi_{\beta} & \sqrt{2}G_{n}\end{array}\right],\]
in which $G_{1},\dots,G_{n}$ are standard Gaussian random variables,
$\chi_{r}$ denotes a chi-distributed random variable with $r$ degrees
of freedom, and all entries in the upper triangular part are independent.

The $\beta=1,2$ cases can be derived by running a tridiagonalization
algorithm on a dense random matrix with density function (\ref{eq:gaussianensembledensity}),
a fact first observed by Trotter \cite{MR0761763}. $H^{\beta}$ is
the natural extension to general $\beta$, and it has the desired
eigenvalue distribution.
\end{defn}
\begin{thm}
[\cite{MR1936554}] For all $\beta>0$, the eigenvalues of the $\beta$-Hermite
matrix model have joint density (\ref{eq:hermitedensity}).
\end{thm}
As $\beta\rightarrow\infty$, the $\beta$-Hermite matrix model converges
in distribution to\[
H^{\infty}=\frac{1}{\sqrt{2}}\left[\begin{array}{cccccc}
0 & \sqrt{n-1}\\
\sqrt{n-1} & 0 & \sqrt{n-2}\\
 & \sqrt{n-2} & 0 & \sqrt{n-3}\\
 &  & \ddots & \ddots & \ddots\\
 &  &  & \sqrt{2} & 0 & \sqrt{1}\\
 &  &  &  & \sqrt{1} & 0\end{array}\right].\]
This matrix encodes the recurrence relation for Hermite polynomials.
In fact, the eigenvalues of this matrix are the roots of the $n$th
polynomial, and the eigenvectors can be expressed easily in terms
of the first $n-1$ polynomials. See \cite{mythesis} and \cite{MR0372517}
for details.

\subsubsection{Laguerre}

The Laguerre ensembles are closely related to Wishart matrices from
multivariate statistics. Just like the Hermite ensembles, the Laguerre
ensembles come in three flavors. The $\beta=1$ flavor is a distribution
on real $m$-by-$n$ matrices. These matrices need not be square,
much less symmetric. The $\beta=2$ flavor is for complex matrices,
and the $\beta=4$ flavor is for quaternion matrices. In all cases,
the density function is\begin{equation}
\text{const}\times\exp\left(-\frac{\beta}{2}\tr A^{*}A\right)dA,\label{eq:wishartdensity}\end{equation}
in which $A^{*}$ denotes the conjugate transpose of $A$.

For a Laguerre matrix with $m$ rows and $n$ columns, let $a=m-n$.
It is well known that the singular values of this matrix are described
by the density\begin{equation}
\text{const}\times e^{-\frac{\beta}{2}\sum_{i=1}^{n}\lambda_{i}}\prod_{i=1}^{n}\lambda_{i}^{\frac{\beta}{2}(a+1)-1}\prod_{1\leq i<j\leq n}|\lambda_{i}-\lambda_{j}|^{\beta}\prod_{i=1}^{n}d\lambda_{i},\label{eq:laguerredensity}\end{equation}
in which $\lambda_{i}$ is the square of the $i$th singular value.
As usual, $\beta=1$ for real entries and $\beta=2$ for complex entries.

Dumitriu and Edelman also extended this family of random matrix distributions
to all $\beta>0$ and nonintegral $a$. The notation in this article
differs from the original notation, instead following \cite{mythesis}.

\begin{defn}
The $n$-by-$n$ \emph{$\beta$-Laguerre matrix model}, parameterized
by $a>-1$, is the distribution on real symmetric matrices\[
L^{\beta,a}\sim\frac{1}{\sqrt{\beta}}\left[\begin{array}{ccccc}
\chi_{(a+n)\beta} & \chi_{(n-1)\beta}\\
 & \chi_{(a+n-1)\beta} & \chi_{(n-2)\beta}\\
 &  & \ddots & \ddots\\
 &  &  & \chi_{(a+2)\beta} & \chi_{\beta}\\
 &  &  &  & \chi_{(a+1)\beta}\end{array}\right].\]
in which $\chi_{r}$ denotes a chi-distributed random variable with
$r$ degrees of freedom, and all entries are independent. The $(n+1)$-by-$n$
\emph{$\beta$-Laguerre matrix model}, parameterized by $a>0$, is\[
M^{\beta,a}\sim\frac{1}{\sqrt{\beta}}\left[\begin{array}{ccccc}
\chi_{n\beta}\\
\chi_{(a+n-1)\beta} & \chi_{(n-1)\beta}\\
 & \chi_{(a+n-2)\beta} & \chi_{(n-2)\beta}\\
 &  & \ddots & \ddots\\
 &  &  & \chi_{(a+1)\beta} & \chi_{\beta}\\
 &  &  &  & \chi_{a\beta}\end{array}\right],\]
with independent entries.
\end{defn}
Notice that $(L^{\beta,a})(L^{\beta,a})^{T}$ and $(M^{\beta,a})^{T}(M^{\beta,a})$
are identically distributed symmetric tridiagonal matrices. This tridiagonal
matrix is actually what Dumitriu and Edelman termed the $\beta$-Laguerre
matrix model. For more information on why we consider two different
random bidiagonal matrices, see  \cite{mythesis}.

The $\beta=1,2$ cases can be derived from dense random matrices following
the density (\ref{eq:wishartdensity}), via a bidiagonalization algorithm,
a fact first observed by Silverstein \cite{MR0806232}. Then the general
$\beta$ matrix model is obtained by extending in the natural way.

\begin{thm}
[\cite{MR1936554}]For all $\beta>0$ and $a>-1$, the singular
values, squared, of $L^{\beta,a}$ follow the density (\ref{eq:laguerredensity}).
For all $\beta>0$ and $a>0$, the singular values, squared, of $M^{\beta,a}$
follow the density (\ref{eq:laguerredensity}).
\end{thm}
As $\beta\rightarrow\infty$, the $n$-by-$n$ $\beta$-Laguerre matrix
model approaches\begin{equation}
L^{\infty,a}=\left[\begin{array}{ccccc}
\sqrt{a+n} & \sqrt{n-1}\\
 & \sqrt{a+n-1} & \sqrt{n-2}\\
 &  & \ddots & \ddots\\
 &  &  & \sqrt{a+2} & \sqrt{1}\\
 &  &  &  & \sqrt{a+1}\end{array}\right]\label{eq:laguerreinf}\end{equation}
in distribution, and the $(n+1)$-by-$n$ $\beta$-Laguerre matrix
model approaches\begin{equation}
M^{\infty,a}=\left[\begin{array}{ccccc}
\sqrt{n}\\
\sqrt{a+n-1} & \sqrt{n-1}\\
 & \sqrt{a+n-2} & \sqrt{n-2}\\
 &  & \ddots & \ddots\\
 &  &  & \sqrt{a+1} & \sqrt{1}\\
 &  &  &  & \sqrt{a}\end{array}\right]\label{eq:laguerrerectangularinf}\end{equation}
in distribution. The nonzero singular values, squared, of both of
these matrices are the roots of the $n$th Laguerre polynomial with
parameter $a$, and the singular vectors are expressible in terms
of the first $n-1$ polynomials. See \cite{mythesis} and \cite{MR0372517}
for details.

\subsubsection{\label{sec:jacobimodel}Jacobi}

Our presentation of the Jacobi matrix model is somewhat unorthodox.
A more detailed exposition can be found in \cite{jacobimodel}.

Consider the space of $(2n+a+b)$-by-$(2n+a+b)$ real orthogonal matrices.
A \emph{CS decomposition} of a matrix $X$ from this distribution
can be computed by partitioning $X$ into rectangular blocks of size
$(n+a)$-by-$n$, $(n+a)$-by-$(n+a+b)$, $(n+b)$-by-$n$, and $(n+b)$-by-$(n+a+b)$,\[
X=\left[\begin{array}{c|c}
X_{11} & X_{12}\\
\hline X_{21} & X_{22}\end{array}\right],\]
and computing singular value decompositions for the four blocks. Because
$X$ is orthogonal, something fortuitous happens: all four blocks
have essentially the same singular values, and there is much sharing
of singular vectors. In fact, $X$ can be factored as\begin{equation}
X=\left[\begin{array}{c|c}
U_{1}\\
\hline  & U_{2}\end{array}\right]\left[\begin{array}{c|c}
C & S\\
\hline -S & C\end{array}\right]\left[\begin{array}{c|c}
V_{1}\\
\hline  & V_{2}\end{array}\right]^{T},\label{eq:csd}\end{equation}
in which $U_{1}$, $U_{2}$, $V_{1}$, and $V_{2}$ are orthogonal
and $C$ and $S$ are nonnegative diagonal. This is the CS decomposition,
and the diagonal entries $c_{1},c_{2},\dots,c_{n}$ of $C$ are knows
as \emph{CS values}. An analogous decomposition exists for complex
unitary matrices $X$, involving unitary $U_{1}$, $U_{2}$, $V_{1}$,
and $V_{2}$.

The Jacobi matrix model is defined by placing Haar measure on $X$.
The resulting distribution on CS values is most conveniently described
in terms of $\lambda_{i}=c_{i}^{2}$, $i=1,\dots,n$, which have joint
density \begin{equation}
\text{const}\times\prod_{i=1}^{n}\lambda_{1}^{\frac{\beta}{2}(a+1)-1}(1-\lambda_{i})^{\frac{\beta}{2}(b+1)-1}\prod_{i<j}|\lambda_{i}-\lambda_{j}|^{\beta}\prod_{i=1}^{n}d\lambda_{i}.\label{eq:jacobidensity}\end{equation}

The Jacobi matrix model has been extended beyond the real and complex
cases ($\beta=1,2$) to general $\beta>0$, first by Killip and Nenciu
and later by the authors of the present article \cite{jacobimodel,killipnenciu,mythesis}.
The following definition involves the beta distribution $\betapdf(c,d)$
on the interval $(0,1)$, whose density function is $\frac{\Gamma(c+d)}{\Gamma(c)\Gamma(d)}x^{c-1}(1-x)^{d-1}$.

\begin{defn}
The $2n$-by-$2n$ \emph{$\beta$-Jacobi matrix model} $J^{\beta,a,b}$,
parameterized by $\beta>0$, $a>-1$, and $b>-1$, is a distribution
on orthogonal matrices with a special structure called bidiagonal
block form. It is defined in terms of random angles $\theta_{1},\dots,\theta_{n}$
and $\phi_{1},\dots,\phi_{n-1}$ from $[0,\frac{\pi}{2}]$. All $2n-1$
angles are independent, and their distributions are defined by\begin{align*}
\cos^{2}\theta_{i} & \sim\betapdf({\textstyle \frac{\beta}{2}}(a+i),{\textstyle \frac{\beta}{2}}(b+i))\\
\cos^{2}\phi_{i} & \sim\betapdf({\textstyle \frac{\beta}{2}}i,{\textstyle \frac{\beta}{2}}(a+b+1+i)).\end{align*}
The entries of the $\beta$-Jacobi matrix model are expressed in terms
of $c_{i}=\cos\theta_{i}$, $s_{i}=\sin\theta_{i}$, $c'_{i}=\cos\phi_{i}$,
and $s'_{i}=\sin\phi_{i}$. \[
J^{\beta,a,b}\sim\left[\begin{array}{cccc|cccc}
c_{n} & -s_{n}c'_{n-1} &  &  & s_{n}s'_{n-1}\\
 & c_{n-1}s'_{n-1} & \ddots &  & c_{n-1}c'_{n-1} & \ddots\\
 &  & \ddots & -s_{2}c'_{1} &  & \ddots & s_{2}s'_{1}\\
 &  &  & c_{1}s'_{1} &  &  & c_{1}c'_{1} & s_{1}\\
\hline -s_{n} & -c_{n}c'_{n-1} &  &  & c_{n}s'_{n-1}\\
 & -s_{n-1}s'_{n-1} & \ddots &  & -s_{n-1}c'_{n-1} & \ddots\\
 &  & \ddots & -c_{2}c'_{1} &  & \ddots & c_{2}s'_{1}\\
 &  &  & -s_{1}s'_{1} &  &  & -s_{1}c'_{1} & c_{1}\end{array}\right].\]

\end{defn}
\begin{thm}
[\cite{jacobimodel,killipnenciu}] Partition the $2n$-by-$2n$
$\beta$-Jacobi matrix model into four blocks of size $n$-by-$n$.
The resulting CS values, squared, have density function (\ref{eq:jacobidensity}).
This is true for all $\beta>0$.
\end{thm}
As $\beta\rightarrow\infty$, the angles $\theta_{1},\dots,\theta_{n}$
and $\phi_{1},\dots,\phi_{n-1}$ converge in distribution to deterministic
angles $\bar{\theta}_{1},\dots,\bar{\theta}_{n}$ and $\bar{\phi}_{1},\dots,\bar{\phi}_{n-1}$,
whose cosines and sines will be denoted $\bar{c}_{i}$, $\bar{s}_{i}$,
$\bar{c}'_{i}$, and $\bar{s}'_{i}$. It is not difficult to show
$\bar{c}_{i}=\sqrt{\frac{a+i}{a+b+2i}}$ and $\bar{c}'_{i}=\sqrt{\frac{i}{a+b+1+2i}}$.
Because the angles have deterministic limits, the matrix model itself
converges in distribution to a fixed matrix $J^{\infty,a,b}$. The
entries of $J^{\infty,a,b}$ encode the recurrence relation for Jacobi
polynomials. The CS values, squared, of $J^{\infty,a,b}$ are the
roots of the $n$th Jacobi polynomial with parameters $a,b$, and
the entries of $U_{1}$, $U_{2}$, $V_{1}$, and $V_{2}$ are expressible
in terms of the first $n-1$ polynomials. See \cite{mythesis} and
\cite{MR0372517} for details.

\subsection{Local eigenvalue behavior}

The three classes of local behavior indicated in Figure \ref{fig:softhardbulk}
are observed by taking $n\rightarrow\infty$ limits of random matrices,
carefully translating and rescaling along the way to focus on a particular
region of the spectrum. This section records the constants required
in some interesting rescalings.

For references concerning the material below, consult the introduction
to this article. Note that much of the existing theory, including
many results concerning the existence of large $n$ limiting distributions
and explicit formulas for those distributions, is restricted to the
cases $\beta=1,2,4$. Further progress in general $\beta$ random
matrix theory may be needed before discussion of general $\beta$
distributions is perfectly well founded. Although these technical
issues are certainly important in the context of the stochastic operator
approach, the concrete results later in this article do not depend
on any subtle probabilistic issues, and hence, we dispense with such
technical issues for the remainder of this section.

\subsubsection{Soft edge}

Soft edge behavior, suggested by Figure \ref{fig:softhardbulk}(a),
can be seen in the right (and left) edge of the Hermite spectrum and
the right edge of the Laguerre spectrum.

In the case of Hermite, the $k$th largest eigenvalue $\lambda_{n+1-k}(H^{\beta})$
displays soft edge behavior. Specifically, $-\sqrt{2}n^{1/6}(\lambda_{n+1-k}(H^{\beta})-\sqrt{2n})$
approaches a soft edge distribution as $n\rightarrow\infty$.

In the case of Laguerre, the $k$th largest singular value $\sigma_{n+1-k}(L^{\beta,a})$
displays soft edge behavior. Specifically, $-2^{2/3}n^{1/6}(\sigma_{n+1-k}(L^{\beta,a})-2\sqrt{n})$
approaches a soft edge distribution as $n\rightarrow\infty$. Of course,
the $k$th largest singular value of the rectangular model $M^{\beta,a}$
displays the same behavior, because the nonzero singular values of
the two models have the same joint distribution.

\subsubsection{Hard edge}

Hard edge behavior, suggested by Figure \ref{fig:softhardbulk}(b),
can be seen in the left edge of the Laguerre spectrum and the left
and right edges of the Jacobi spectrum.

In the case of Laguerre, the $k$th smallest singular value $\sigma_{k}(L^{\beta,a})$
displays hard edge behavior. Specifically, $\sqrt{2}\sqrt{2n+a+1}\sigma_{k}(L^{\beta,a})$
approaches a hard edge distribution as $n\rightarrow\infty$. Of course,
the $k$th smallest nonzero singular value of $M^{\beta,a}$ displays
the same behavior, because the singular values of the two matrix models
have the same joint density.

In the case of Jacobi, the $k$th smallest CS value displays hard
edge behavior. Specifically, let $c_{kk}(J^{\beta,a,b})$ be the $k$th
smallest diagonal entry in the matrix $C$ of (\ref{eq:csd}), when
applied to the $\beta$-Jacobi matrix model. Then $(2n+a+b+1)c_{kk}(J^{\beta,a,b})$
approaches a hard edge distribution as $n\rightarrow\infty$.

\subsubsection{Bulk}

Bulk behavior is seen in the interior of spectra, as opposed to the
edges. Suppose that the least and greatest members of the spectrum
of an $n$-by-$n$ random matrix are $O(L_{n})$ and $O(R_{n})$,
respectively, as $n\rightarrow\infty$. Then bulk behavior can often
be seen in the spacings between consecutive eigenvalues near the point
$(1-p)L_{n}+pR_{n}$, for any constant $p\in(0,1)$, as $n\rightarrow\infty$.
This is true for the Hermite, Laguerre, and Jacobi ensembles. Because
this article does not consider bulk spacings in great detail, the
constants involved in the scalings are omitted.

\subsection{\label{section:backgroundfinitediffs}Finite difference schemes}

The solution to a differential equation can be approximated numerically
through a \emph{finite difference scheme}. This procedure works by
replacing various differential operators with matrices that mimic
their behavior. For example, the first derivative operator can be
discretized by a matrix whose action amounts to subtracting function
values at nearby points on the real line, essentially omitting the
limit in the definition of the derivative.

With this in mind, $\nabla_{m,n}$ is defined to be the $m$-by-$n$
upper bidiagonal matrix with $1$ on the superdiagonal and $-1$ on
the main diagonal,\[
\nabla_{m,n}=\left[\begin{array}{cccc}
-1 & 1\\
 & -1 & 1\\
 &  & \ddots & \ddots\end{array}\right].\]
The subscripts are omitted when the size of the matrix is clear from
context. Up to a constant factor, $\nabla_{m,n}$ encodes a finite
difference scheme for the first derivative operator when certain boundary
conditions are in place.

The matrix $\Delta_{n}$ is defined to be the symmetric tridiagonal
matrix with 2 on the main diagonal and -1
 on the superdiagonal and subdiagonal,\[
\Delta_{n}=\left[\begin{array}{ccccc}
2 & -1\\
-1 & 2 & -1\\
 & \ddots & \ddots & \ddots\\
 &  & -1 & 2 & -1\\
 &  &  & -1 & 2\end{array}\right].\]
 Note that $\Delta_{n}=\nabla_{n,n+1}\nabla_{n,n+1}^{T}$. Under certain
conditions, $\Delta_{n}$ discretizes the second derivative operator,
up to a constant factor.

A few other matrices prove useful when constructing finite difference
schemes. $\Omega_{n}$ denotes the $n$-by-$n$ diagonal matrix with
$-1,1,-1,1,\dots$ along the main diagonal. $F_{n}$ denotes the $n$-by-$n$
{}``flip'' permutation matrix, with ones along the diagonal from
top-right to bottom-left. In both cases, the subscript is omitted
when the size of the matrix is clear. Finally, the {}``interpolating
matrix'' $S_{m,n}=-\frac{1}{2}\Omega_{m}\nabla_{m,n}\Omega_{n}$
proves useful when constructing finite difference schemes for which
the domain and codomain meshes interleave. $S_{m,n}$ is the $m$-by-$n$
upper bidiagonal matrix in which every entry on the main diagonal
and superdiagonal equals $\frac{1}{2}$. Subscripts will be omitted
where possible.

\section{\label{sec:results}Results}

This section defines the stochastic Airy and Bessel operators, briefly
mentions the stochastic sine operator, and states results that are
proved in later sections.

\subsection{The stochastic differential operators}

\subsubsection{Stochastic Airy operator}

\begin{defn}
The \emph{classical Airy operator} is\[
\mathcal{A}^{\infty}=-\frac{d^{2}}{dx^{2}}+x,\]
acting on $L^{2}((0,\infty))$ functions $v$ satisfying the boundary
conditions $v(0)=0$, $\lim_{x\rightarrow\infty}v(x)=0$.
\end{defn}
An eigenvalue-eigenfunction pair consists of a number $\lambda$ and
a function $v$ such that $\mathcal{A}^{\infty}v=\lambda v$. The
complete eigenvalue decomposition is\[
(\mathcal{A}^{\infty}-\lambda_{k})\Ai(-x-\lambda_{k})=0,\]
for $k=1,2,3,\dots$, in which $\Ai$ denotes the unique solution
to Airy's equation $f''(x)=xf(x)$ that decays as $x\rightarrow\infty$,
and $\lambda_{k}$ equals the negation of the $k$th zero of $\Ai$.
As typical with Sturm-Liouville operators, $\mathcal{A}^{\infty}$
acts naturally on a subspace of Sobolev space and can be extended
to all $L^{2}((0,\infty))$ functions satisfying the boundary conditions
via the eigenvalue decomposition.

Intuitively, the \emph{stochastic} Airy operator is obtained by adding
white noise, the formal derivative of Brownian motion,\begin{equation}
-\frac{d^{2}}{dx^{2}}+x+\frac{2}{\sqrt{\beta}}B'(x).\label{eq:intuitiveairy}\end{equation}
However, white noise sometimes poses technical difficulties. To avoid
these potential difficulties, we express the stochastic Airy operator
in terms of a conjugation of a seemingly simpler operator, i.e., by
changing variables. 

\begin{defn}
\label{def:stochasticairy}Let $\beta>0$, let $B(x)$ be a Brownian
path on $(0,\infty)$, and let \begin{equation}
\phi(x)=\exp\left(\frac{2}{\sqrt{\beta}}\int_{0}^{x}B(x)dx\right).\label{eq:phi}\end{equation}
The \emph{stochastic Airy operator} $\mathcal{A}^{\beta}$ acts on
functions $v(x)=f(x)\phi(x)$ satisfying the boundary conditions $v(0)=0$,
$\lim_{x\rightarrow\infty}v(x)=0$. It is defined by\[
[\mathcal{A}^{\beta}(f\phi)](x)=\phi(x)\cdot\left([\mathcal{A}^{\infty}f](x)-\frac{4}{\sqrt{\beta}}B(x)f'(x)-\frac{4}{\beta}B(x)^{2}f(x)\right),\]
or, to abbreviate,\begin{equation}
\mathcal{A}^{\beta}=\phi\left(\mathcal{A}^{\infty}-\frac{4}{\sqrt{\beta}}B\frac{d}{dx}-\frac{4}{\beta}B^{2}\right)\phi^{-1}.\label{eq:technicalairy}\end{equation}

\end{defn}
An eigenvalue-eigenfunction pair consists of a real number $\lambda$
and a function $v$ such that $\mathcal{A}^{\beta}v=\lambda v$.

Note that $\phi(x)$ is defined in terms of a Riemann integral of
Brownian motion, which is continuous. This is not a stochastic integral,
and nothing like an It\^{o} or Stratonovich interpretation must be
specified.

When $\beta=\infty$, the stochastic Airy operator equals the classical
Airy operator. When $\beta<\infty$, equations (\ref{eq:intuitiveairy})
and (\ref{eq:technicalairy}) are formally equivalent. To see this,
apply $\mathcal{A}^{\beta}$ to $v=f\phi$, proceeding formally as
follows. Combining\begin{multline*}
\phi\mathcal{A}^{\infty}(\phi^{-1}v)=-\phi(\phi^{-1})''v-2\phi(\phi^{-1})'v'-\phi\phi^{-1}v''+xv\\
=\frac{2}{\sqrt{\beta}}B'v-\frac{4}{\beta}B^{2}v+\frac{4}{\sqrt{\beta}}Bv'+\mathcal{A}^{\infty}v\end{multline*}
and\begin{multline*}
\phi\cdot\left(-\frac{4}{\sqrt{\beta}}B\frac{d}{dx}\right)(\phi^{-1}v)=-\frac{4}{\sqrt{\beta}}\phi B(\phi^{-1})'v-\frac{4}{\sqrt{\beta}}\phi B\phi^{-1}v'\\
=\frac{8}{\beta}B^{2}v-\frac{4}{\sqrt{\beta}}Bv'\end{multline*}
and\begin{align*}
\phi\cdot\left(-\frac{4}{\beta}B^{2}\right)(\phi^{-1}v) & =-\frac{4}{\beta}B^{2}v\end{align*}
yields\begin{align*}
\mathcal{A}^{\beta}v & =\mathcal{A}^{\infty}v+\frac{2}{\sqrt{\beta}}B'v.\end{align*}

The stochastic Airy operator acts naturally on any function of the
form $f\phi$, in which $f$ has two derivatives. Also, the Rayleigh
quotient defined by $\mathcal{A}^{\beta}$, \[
\frac{\langle v,\mathcal{A}^{\beta}v\rangle}{\langle v,v\rangle}=\frac{\int(v')^{2}dx+\int xv^{2}dx+\frac{2}{\sqrt{\beta}}\int v^{2}dB}{\int v^{2}dx},\]
is well defined and does not require an It\^{o} or Stratonovich interpretation
if $v$ is deterministic, decays sufficiently fast, and is sufficiently
smooth, say, if it has a bounded first derivative. See \cite{oksendal}.

\subsubsection{Stochastic Bessel operator}

\begin{defn}
The \emph{classical Bessel operator with type (i) boundary conditions},
parameterized by $a>-1$, is the operator whose action is\[
\mathcal{J}_{a}^{\infty}=-2\sqrt{x}\frac{d}{dx}+\frac{a}{\sqrt{x}},\]
acting on functions $v$ satisfying\begin{equation}
\text{type (i) b.c.'s: }v(1)=0\text{ and }(\mathcal{J}_{a}^{\infty}v)(0)=0.\label{eq:typeibcs}\end{equation}
We will abuse notation and also denote the \emph{classical Bessel
operator with type (ii) boundary conditions} by $\mathcal{J}_{a}^{\infty}$.
The action of this operator is also $\mathcal{J}_{a}^{\infty}=-2\sqrt{x}\frac{d}{dx}+\frac{a}{\sqrt{x}}$,
and it is defined for all $a>-1$, but its domain consists of functions
$v$ satisfying\begin{equation}
\text{type (ii) b.c.'s: }(\mathcal{J}_{a}^{\infty}v)(0)=0\text{ and }(\mathcal{J}_{a}^{\infty}v)(1)=0.\label{eq:typeiibcs}\end{equation}

\end{defn}
The adjoint of the classical Bessel operator (with either type (i)
or type (ii) boundary conditions) has action\[
(\mathcal{J}_{a}^{\infty})^{*}=2\sqrt{x}\frac{d}{dx}+\frac{a+1}{\sqrt{x}}.\]
The singular value decompositions are defined in terms of the Bessel
functions of the first kind $j_{a}$ by\begin{align*}
\text{type (i) b.c.'s: } & \mathcal{J}_{a}^{\infty}[j_{a}(\sigma_{k}\sqrt{x})]=\sigma_{k}j_{a+1}(\sigma_{k}\sqrt{x})\\
 & (\mathcal{J}_{a}^{\infty})^{*}[j_{a+1}(\sigma_{k}\sqrt{x})]=\sigma_{k}j_{a}(\sigma_{k}\sqrt{x})\\
 & 0<\sigma_{1}<\sigma_{2}<\sigma_{3}<\cdots\text{ zeros of $j_{a}$}\end{align*}
and \begin{align*}
\text{type (ii) b.c.'s: } & \mathcal{J}_{a}^{\infty}[x^{a/2}]=0\\
 & \mathcal{J}_{a}^{\infty}[j_{a}(\sigma_{k}\sqrt{x})]=\sigma_{k}j_{a+1}(\sigma_{k}\sqrt{x})\\
 & (\mathcal{J}_{a}^{\infty})^{*}[j_{a+1}(\sigma_{k}\sqrt{x})]=\sigma_{k}j_{a}(\sigma_{k}\sqrt{x})\\
 & 0<\sigma_{1}<\sigma_{2}<\sigma_{3}<\cdots\text{ zeros of $j_{a+1}$.}\end{align*}
The purposes of the boundary conditions are now clear. The condition
at $x=1$ produces a discrete spectrum, and the condition at $x=0$
eliminates Bessel functions of the second kind, leaving left singular
functions that are nonsingular at the origin.

Intuitively, the \emph{stochastic} Bessel operator is obtained by
adding white noise to obtain \begin{equation}
\mathcal{J}_{a}^{\beta}=\mathcal{J}_{a}^{\infty}+\frac{2}{\sqrt{\beta}}B',\label{eq:intuitivebessel}\end{equation}
with adjoint $(\mathcal{J}_{a}^{\beta})^{*}=(\mathcal{J}_{a}^{\infty})^{*}+\frac{2}{\sqrt{\beta}}B'$.
However, the following definition, which avoids the language of white
noise, offers certain technical advantages.

\begin{defn}
\label{def:stochasticbessel}Let $a>-1$ and $\beta>0$, let $B(x)$
be a Brownian path on $(0,1)$, and let \begin{equation}
\psi(x)=\exp\left(-\frac{1}{\sqrt{\beta}}\int_{x}^{1}w^{-1/2}dB(w)\right).\label{eq:psi}\end{equation}
The \emph{stochastic Bessel operator}, denoted $\mathcal{J}_{a}^{\beta}$,
has action\[
[\mathcal{J}_{a}^{\beta}(f\psi)](x)=\psi(x)\cdot[\mathcal{J}_{a}^{\infty}f](x),\]
or, to abbreviate,\begin{equation}
\mathcal{J}_{a}^{\beta}=\psi\mathcal{J}_{a}^{\infty}\psi^{-1}.\label{eq:technicalbessel}\end{equation}
Either type (i) or type (ii) boundary conditions may be applied. (See
(\ref{eq:typeibcs}) and (\ref{eq:typeiibcs}).)
\end{defn}
The adjoint is\[
(\mathcal{J}_{a}^{\beta})^{*}=\psi^{-1}(\mathcal{J}_{a}^{\infty})^{*}\psi.\]

The function $\psi$ involves a stochastic integral, but because the
integrand is smooth and not random, it is not necessary to specify
an It\^{o} or Stratonovich interpretation.

Note that when $\beta=\infty$, the stochastic Bessel operator equals
the classical Bessel operator. When $\beta<\infty$, equations (\ref{eq:intuitivebessel})
and (\ref{eq:technicalbessel}) are formally equivalent. To see this,
apply $\mathcal{J}_{a}^{\beta}$ to $v=f\phi$, proceeding formally
as follows.\begin{multline*}
\mathcal{J}_{a}^{\beta}v=\psi\left(-2\sqrt{x}(\psi^{-1})'v-2\sqrt{x}\psi^{-1}v'+\frac{a}{\sqrt{x}}\psi^{-1}v\right)\\
=-2\sqrt{x}\frac{(\psi^{-1})'}{\psi^{-1}}v-2\sqrt{x}v'+\frac{a}{\sqrt{x}}v\\
=-2\sqrt{x}\left(-\frac{1}{\sqrt{\beta}}x^{-1/2}B'\right)v-2\sqrt{x}v'+\frac{a}{\sqrt{x}}v=\mathcal{J}_{a}^{\infty}v+\frac{2}{\sqrt{\beta}}B'v.\end{multline*}

The stochastic Bessel operator acts naturally on any function of the
form $f\psi$ for which $f$ has one derivative, assuming the boundary
conditions are satisfied. Its adjoint acts naturally on functions
of the form $g\psi^{-1}$ for which $g$ has one derivative and the
boundary conditions are satisfied.

Sometimes, expressing the stochastic Bessel operator in Liouville
normal form proves to be useful. The classical Bessel operator in
Liouville normal form, denoted $\tilde{\mathcal{J}}_{a}^{\infty}$,
is defined by\begin{align*}
\tilde{\mathcal{J}}_{a}^{\infty} & =-\frac{d}{dx}+(a+{\textstyle \frac{1}{2}})\frac{1}{x}, & (\tilde{\mathcal{J}}_{a}^{\infty})^{*} & =\frac{d}{dx}+(a+{\textstyle \frac{1}{2}})\frac{1}{x},\end{align*}
with either type (i) or type (ii) boundary conditions. The singular
values remain unchanged, while the singular functions undergo a change
of variables. The SVD's are\begin{align}
\text{type (i) b.c.'s:}\; & \mathcal{\tilde{J}}_{a}^{\infty}[\sqrt{x}j_{a}(\sigma_{k}x)]=\sigma_{k}\sqrt{x}j_{a+1}(\sigma_{k}x)\label{eq:besselliouvillesvd}\\
 & (\mathcal{\tilde{J}}_{a}^{\infty})^{*}[\sqrt{x}j_{a+1}(\sigma_{k}x)]=\sigma_{k}\sqrt{x}j_{a}(\sigma_{k}x)\nonumber \\
 & 0<\sigma_{1}<\sigma_{2}<\sigma_{3}<\cdots\text{ zeros of $j_{a}$}\nonumber \end{align}
and \begin{align}
\text{type (ii) b.c.'s:}\; & \mathcal{\tilde{J}}_{a}^{\infty}[x^{a+1/2}]=0\\
 & \mathcal{\tilde{J}}_{a}^{\infty}[\sqrt{x}j_{a}(\sigma_{k}x)]=\sigma_{k}\sqrt{x}j_{a+1}(\sigma_{k}x)\nonumber \\
 & (\mathcal{\tilde{J}}_{a}^{\infty})^{*}[\sqrt{x}j_{a+1}(\sigma_{k}x)]=\sigma_{k}\sqrt{x}j_{a}(\sigma_{k}x)\nonumber \\
 & 0<\sigma_{1}<\sigma_{2}<\sigma_{3}<\cdots\text{ zeros of $j_{a+1}$.}\nonumber \end{align}
Note that although the change of variables to Liouville normal form
affects asymptotics near 0, the original boundary conditions still
serve their purposes.

\begin{defn}
Let $a>-1$ and $\beta>0$, and let $B(x)$ be a Brownian path on
$(0,1)$. The \emph{stochastic Bessel operator in Liouville normal
form}, denoted $\mathcal{\tilde{J}}_{a}^{\beta}$, has action\[
[\tilde{\mathcal{J}}_{a}^{\beta}(f\psi^{\sqrt{2}})](x)=\psi(x)^{\sqrt{2}}\cdot[\tilde{\mathcal{J}}_{a}^{\infty}f](x),\]
or, to abbreviate,\[
\tilde{\mathcal{J}}_{a}^{\beta}=\psi^{\sqrt{2}}\tilde{\mathcal{J}}_{a}^{\infty}\psi^{-\sqrt{2}}.\]
$\psi$ is defined in (\ref{eq:psi}). Either type (i) or type (ii)
boundary conditions may be applied.

This operator acts naturally on functions of the form $f\psi^{\sqrt{2}}$
for which $f$ is once differentiable. It is formally equivalent to
$\tilde{\mathcal{J}}_{a}^{\infty}+\sqrt{\frac{2}{\beta}}\frac{1}{\sqrt{x}}B$:\begin{multline*}
\tilde{\mathcal{J}}_{a}^{\beta}v=\psi^{\sqrt{2}}\left(\sqrt{2}\psi^{-\sqrt{2}-1}\psi'v-\psi^{-\sqrt{2}}v'+(a+{\textstyle \frac{1}{2}})\frac{1}{x}\psi^{-\sqrt{2}}v\right)\\
=\sqrt{2}\frac{\psi'}{\psi}v-v'+(a+{\textstyle \frac{1}{2}})\frac{1}{x}v=\tilde{\mathcal{J}}_{a}^{\infty}v+\sqrt{\frac{2}{\beta}}\frac{1}{\sqrt{x}}B'v.\end{multline*}

\end{defn}

\subsubsection{Stochastic sine operator}

The last section of this article presents some ideas concerning a
third stochastic differential operator, the \emph{stochastic sine
operator}. This operator likely has the form\[
\left[\begin{array}{c|c}
 & -\frac{d}{dx}\\
\hline \frac{d}{dx}\end{array}\right]+\left[\begin{array}{c|c}
\text{``noise''} & \text{``noise''}\\
\hline \text{``noise''} & \text{``noise''}\end{array}\right].\]
 Key to understanding the stochastic Airy and Bessel operators are
the changes of variables, in terms of $\phi$ and $\psi$, respectively,
that replace white noise with Brownian motion. No analogous change
of variables has yet been found for the stochastic sine operator,
so most discussion of this operator will be left for a future article.

\subsection{\label{sec:discretizationclaims}Random matrices discretize stochastic
differential operators}

Much of the remainder of the article is devoted to supporting the
following claims, relating the stochastic differential operators of
the previous section to the classical ensembles of random matrix theory.
The claims involve {}``scaling'' random matrix models. This is explained
in Sections \ref{sec:zerotempasfinitediff} and \ref{sec:randomasfinitediff}.

\subsubsection{Soft edge $\leftrightarrow$ stochastic Airy operator}

\begin{claim}
\label{claim:hermitesoftfinitediff}The Hermite matrix model, scaled
at the soft edge, encodes a finite difference scheme for the stochastic
Airy operator. See Theorems \ref{thm:infhermitetoairy} and \ref{thm:betahermitetoairy}.
\end{claim}
{}

\begin{claim}
\label{claim:laguerresoftfinitediff}The Laguerre matrix models, scaled
at the soft edge, encode finite difference schemes for the stochastic
Airy operator. See Theorems \ref{thm:inflaguerretoairy} and \ref{thm:betalaguerretoairy}.
\end{claim}

\subsubsection{Hard edge $\leftrightarrow$ stochastic Bessel operator}

\begin{claim}
\label{claim:laguerrehardfinitediff}The Laguerre matrix models, scaled
at the hard edge, encode finite difference schemes for the stochastic
Bessel operator. $L_{\text{hard}}^{\beta,a}$ encodes type (i) boundary
conditions, and $M_{\text{hard}}^{\beta,a}$ encodes type (ii) boundary
conditions. See Theorems \ref{thm:inflaguerreltobessel}, \ref{thm:inflaguerremtobessel},
\ref{thm:betalaguerretobessel}, and \ref{thm:betalaguerremtobessel}.
\end{claim}
{}

\begin{claim}
\label{claim:jacobihardfinitediff}The Jacobi matrix model, scaled
at the hard edge, encodes a finite difference scheme for the stochastic
Bessel operator in Liouville normal form with type (i) boundary conditions.
See Theorems \ref{thm:infjacobitobessel} and \ref{thm:betajacobitobessel}.
\end{claim}

\subsection{Eigenvalues/singular values of stochastic differential operators}

Based on the claims in the previous section, we propose distributions
for the eigenvalues of the stochastic Airy operator and the singular
values of the stochastic Bessel operators.

\begin{conjecture}
The $k$th least eigenvalue of the stochastic Airy operator follows
the $k$th soft edge distribution, with the same value for $\beta$.
\end{conjecture}
The conjecture now appears to be a theorem, due to a proof of Ram\'{i}rez,
Rider, and Vir\'{a}g \cite{rideretal}.

{}

\begin{conjecture}
The $k$th least singular value of the stochastic Bessel operator
with type (i) boundary conditions follows the $k$th hard edge distribution,
with the same values for $\beta$ and $a$. With type (ii) boundary
conditions, the hard edge distribution has parameters $\beta$ and
$a+1$.
\end{conjecture}
The conjecture should be true for both $\mathcal{J}_{a}^{\beta}$
and $\tilde{\mathcal{J}}_{a}^{\beta}$.

\section{Some matrix model identities}

This section establishes relations between various random matrices
which will be useful later in the article. The identities are organized
according to their later application. This section may be skipped
on a first reading.

\subsection{Identities needed for the soft edge}

\begin{lem}
Let $H^{\beta}=(h_{ij})$ be a matrix from the $n$-by-$n$ $\beta$-Hermite
matrix model. $\beta$ may be either finite or infinite. Let $D$
be the diagonal matrix whose $(i,i)$ entry is $(n/2)^{-(i-1)/2}\prod_{k=1}^{i-1}h_{k,k+1}$.
Then if $\beta<\infty$, $DH^{\beta}D^{-1}$ has distribution\begin{equation}
\frac{1}{\sqrt{2\beta}}\left[\begin{array}{ccccc}
\sqrt{2}G_{1} & \sqrt{\beta n}\\
\frac{1}{\sqrt{\beta n}}\chi_{(n-1)\beta}^{2} & \sqrt{2}G_{2} & \sqrt{\beta n}\\
 & \ddots & \ddots & \ddots\\
 &  & \frac{1}{\sqrt{\beta n}}\chi_{2\beta}^{2} & \sqrt{2}G_{n-1} & \sqrt{\beta n}\\
 &  &  & \frac{1}{\sqrt{\beta n}}\chi_{\beta}^{2} & \sqrt{2}G_{n}\end{array}\right],\label{eq:hermiteaftersimilarity}\end{equation}
with all entries independent, while if $\beta=\infty$, $DH^{\infty}D^{-1}$
equals\begin{equation}
\frac{1}{\sqrt{2}}\left[\begin{array}{cccccc}
0 & \sqrt{n}\\
\frac{n-1}{\sqrt{n}} & 0 & \sqrt{n}\\
 & \frac{n-2}{\sqrt{n}} & 0 & \sqrt{n}\\
 &  & \ddots & \ddots & \ddots\\
 &  &  & \frac{2}{\sqrt{n}} & 0 & \sqrt{n}\\
 &  &  &  & \frac{1}{\sqrt{n}} & 0\end{array}\right].\label{eq:infhermiteaftersimilarity}\end{equation}

\end{lem}
{}

\begin{lem}
\label{lem:identityforlairy}Let $L^{\beta,a}=(l_{ij})$ be a matrix
from the $n$-by-$n$ $\beta$-Laguerre matrix model. $\beta$ may
be finite or infinite. Let $D$ be the $2n$-by-$2n$ diagonal matrix
whose $(i,i)$ entry is $n^{-(i-1)/2}\prod_{k=1}^{\lfloor i/2\rfloor}l_{kk}\prod_{k=1}^{\lfloor(i-1)/2\rfloor}l_{k,k+1}$,
and let $P=(p_{ij})$ be the $2n$-by-$2n$ {}``perfect shuffle''
permutation matrix,\[
p_{ij}=\begin{cases}
1 & j=2i-1\text{ or }j=2(i-n)\\
0 & \text{otherwise}\end{cases}.\]
Then, if $\beta<\infty$,\begin{equation}
DP^{T}\left[\begin{array}{cc}
0 & (L^{\beta,a})^{T}\\
L^{\beta,a} & 0\end{array}\right]PD_{L}^{-1}\label{eq:shuffledl}\end{equation}
is the $2n$-by-$2n$ random matrix with independent entries\[
\frac{1}{\sqrt{\beta}}\left[\begin{array}{ccccc}
0 & \sqrt{\beta n}\\
\frac{1}{\sqrt{\beta n}}\chi_{(a+n)\beta}^{2} & 0 & \sqrt{\beta n}\\
 & \frac{1}{\sqrt{\beta n}}\chi_{(n-1)\beta}^{2} & 0 & \sqrt{\beta n}\\
 &  & \frac{1}{\sqrt{\beta n}}\chi_{(a+n-1)\beta}^{2} & 0 & \ddots\\
 &  &  & \frac{1}{\sqrt{\beta n}}\chi_{(n-2)\beta}^{2} & \ddots\\
 &  &  &  & \ddots\end{array}\right],\]
 and if $\beta=\infty$, the matrix of (\ref{eq:shuffledl}) equals
the $2n$-by-$2n$ matrix\[
\left[\begin{array}{ccccc}
0 & \sqrt{n}\\
\frac{a+n}{\sqrt{n}} & 0 & \sqrt{n}\\
 & \frac{n-1}{\sqrt{n}} & 0 & \sqrt{n}\\
 &  & \frac{a+n-1}{\sqrt{n}} & 0 & \ddots\\
 &  &  & \frac{n-2}{\sqrt{n}} & \ddots\\
 &  &  &  & \ddots\end{array}\right].\]

\end{lem}
{}

\begin{lem}
\label{lem:identityformairy}Let $M^{\beta,a}=(m_{ij})$ be a matrix
from the $(n+1)$-by-$n$ $\beta$-Laguerre matrix model. $\beta$
may be finite or infinite. Let $D$ be the diagonal matrix whose $(i,i)$
entry equals $n^{-(i-1)/2}\prod_{k=1}^{\lfloor i/2\rfloor}m_{kk}\prod_{k=1}^{\lfloor(i-1)/2\rfloor}m_{k+1,k}$,
and let $P=(p_{ij})$ be the $(2n+1)$-by-$(2n+1)$ perfect shuffle
matrix,\[
p_{ij}=\begin{cases}
1 & j=2i-1\text{ or }j=2(i-(n+1))\\
0 & \text{otherwise}\end{cases}.\]
 Then, if $\beta<\infty$,\begin{equation}
DP^{T}\left[\begin{array}{cc}
0 & M^{\beta,a}\\
(M^{\beta,a})^{T} & 0\end{array}\right]PD^{-1}\label{eq:shuffledm}\end{equation}
is the $(2n+1)$-by-$(2n+1)$ random matrix with independent entries\[
\frac{1}{\sqrt{\beta}}\left[\begin{array}{ccccc}
0 & \sqrt{\beta n}\\
\frac{1}{\sqrt{\beta n}}\chi_{n\beta}^{2} & 0 & \sqrt{\beta n}\\
 & \frac{1}{\sqrt{\beta n}}\chi_{(a+n-1)\beta}^{2} & 0 & \sqrt{\beta n}\\
 &  & \frac{1}{\sqrt{\beta n}}\chi_{(n-1)\beta}^{2} & 0 & \ddots\\
 &  &  & \frac{1}{\sqrt{\beta n}}\chi_{(a+n-2)\beta}^{2} & \ddots\\
 &  &  &  & \ddots\end{array}\right],\]
with all entries independent, and if $\beta=\infty$, the matrix of
(\ref{eq:shuffledm}) equals the $(2n+1)$-by-$(2n+1)$ matrix\[
\left[\begin{array}{ccccc}
0 & \sqrt{n}\\
\frac{n}{\sqrt{n}} & 0 & \sqrt{n}\\
 & \frac{a+n-1}{\sqrt{n}} & 0 & \sqrt{n}\\
 &  & \frac{n-1}{\sqrt{n}} & 0 & \ddots\\
 &  &  & \frac{a+n-2}{\sqrt{n}} & \ddots\\
 &  &  &  & \ddots\end{array}\right].\]

\end{lem}

\subsection{Identities needed for the hard edge}

The remaining identities are derived from the following two completely
trivial lemmas. The first operates on square bidiagonal matrices,
and the second operates on rectangular bidiagonal matrices.

\begin{lem}
Let $A=(a_{ij})$ and $B=(b_{ij})$ be $n$-by-$n$ upper bidiagonal
matrices with the same sign pattern and with no zero entries on the
main diagonal or superdiagonal. Set\begin{align*}
g_{2i-1} & =-\log|a_{ii}|+\log|b_{ii}|,\; i=1,\dots,n\\
g_{2i} & =\log|a_{i,i+1}|-\log|b_{i,i+1}|,\; i=1,\dots,n-1,\end{align*}
and, for $i=1,\dots,2n$, let $d_{i}=\sum_{k=i}^{2n-1}g_{k}$. Then\[
A=e^{D_{\text{even}}}Be^{-D_{\text{odd}}},\]
with $D_{\text{even}}=\diag(d_{2},d_{4},\dots,d_{2n})$ and $D_{\text{odd}}=\diag(d_{1},d_{3},\dots,d_{2n-1})$.
\end{lem}
\begin{proof}
The $(i,i)$ entry of $e^{D_{\text{even}}}Be^{-D_{\text{odd}}}$ equals
$e^{d_{2i}}b_{ii}e^{-d_{2i-1}}=b_{ii}e^{-g_{2i-1}}=b_{ii}\frac{|a_{ii}|}{|b_{ii}|}=a_{ii}$.
The $(i,i+1)$ entry equals $e^{d_{2i}}b_{i,i+1}e^{-d_{2i+1}}=b_{i,i+1}e^{g_{2i}}=b_{i,i+1}\frac{|a_{i,i+1}|}{|b_{i,i+1}|}=a_{i,i+1}$.
All other entries are zero.
\end{proof}
\begin{lem}
Let $A=(a_{ij})$ and $B=(b_{ij})$ be $n$-by-$(n+1)$ upper bidiagonal
matrices with the same sign pattern and with no zero entries on the
main diagonal or superdiagonal. For $i=1,\dots,n$, set $g_{2i-1}=-\log|a_{ii}|+\log|b_{ii}|$
and $g_{2i}=\log|a_{i,i+1}|-\log|b_{i,i+1}|$, and for $i=1,\dots,2n+1$,
let $d_{i}=\sum_{k=i}^{2n}g_{k}$. Then $A=e^{D_{\text{even}}}Be^{-D_{\text{odd}}}$,
with $D_{\text{even}}=\diag(d_{2},d_{4},\dots,d_{2n})$ and $D_{\text{odd}}=\diag(d_{1},d_{3},\dots,d_{2n+1})$.
\end{lem}
\begin{proof}
The $(i,i)$ entry of $e^{D_{\text{even}}}Be^{-D_{\text{odd}}}$ equals
$e^{d_{2i}}b_{ii}e^{-d_{2i-1}}=b_{ii}e^{-g_{2i-1}}=b_{ii}\frac{|a_{ii}|}{|b_{ii}|}=a_{ii}$.
The $(i,i+1)$ entry of $e^{D_{\text{even}}}Be^{-D_{\text{odd}}}$
equals $e^{d_{2i}}b_{i,i+1}e^{-d_{2i+1}}=b_{i,i+1}e^{g_{2i}}=b_{i,i+1}\frac{|a_{i,i+1}|}{|b_{i,i+1}|}=a_{i,i+1}$.
All other entries are zero.
\end{proof}
The lemmas immediately establish the following three identities. Consult
Section \ref{section:backgroundfinitediffs} for the definitions of
$\Omega$ and $F$.

For the square $\beta$-Laguerre matrix model, we have\begin{multline}
\sqrt{\frac{2}{(2n+a+1)^{-1}}}F\Omega(L^{\beta,a})^{T}\Omega F\label{eq:identityforlhard}\\
\sim e^{D_{\text{even}}}\left(\sqrt{\frac{2}{(2n+a+1)^{-1}}}F\Omega(L^{\infty,a})^{T}\Omega F\right)e^{-D_{\text{odd}}}\end{multline}
with\begin{align*}
D_{\text{even}} & =\diag(d_{2},d_{4},\dots,d_{2n}),\; D_{\text{odd}}=\diag(d_{1},d_{3},\dots,d_{2n-1})\\
d_{i} & ={\textstyle \sum_{k=i}^{2n-1}g_{k}}\\
g_{2i-1} & \sim-\log({\textstyle \frac{1}{\sqrt{\beta}}}\chi_{(a+i)\beta})+\log\sqrt{a+i}\\
g_{2i} & \sim\log({\textstyle \frac{1}{\sqrt{\beta}}}\chi_{i\beta})-\log\sqrt{i}.\end{align*}
 $g_{1},\dots,g_{2n-1}$ are independent.

For the rectangular $\beta$-Laguerre matrix model, we have\begin{multline}
\sqrt{\frac{2}{(2n+a+1)^{-1}}}F_{n}\Omega_{n}(M^{\beta,a})^{T}\Omega_{n+1}F_{n+1}\\
\sim e^{D_{\text{even}}}\left(\sqrt{\frac{2}{(2n+a+1)^{-1}}}F_{n}\Omega_{n}(M^{\infty,a})^{T}\Omega_{n+1}F_{n+1}\right)e^{-D_{\text{odd}}}\label{eq:identityformhard}\end{multline}
with\begin{align*}
D_{\text{even}} & =\diag(d_{2},d_{4},\dots,d_{2n}),\; D_{\text{odd}}=\diag(d_{1},d_{3},\dots,d_{2n+1})\\
d_{i} & ={\textstyle \sum_{k=i}^{2n}g_{k}}\\
g_{2i-1} & \sim-\log({\textstyle \frac{1}{\sqrt{\beta}}}\chi_{(a+i-1)\beta})+\log\sqrt{a+i-1}\\
g_{2i} & \sim\log({\textstyle \frac{1}{\sqrt{\beta}}}\chi_{i\beta})-\log\sqrt{i}.\end{align*}
 $g_{1},\dots,g_{2n}$ are independent.

Finally, for the bottom-right block $B_{22}^{\beta,a,b}$ of the $\beta$-Jacobi
matrix model $J^{\beta,a,b}$, we have\begin{multline}
\frac{1}{(2n+a+b+1)^{-1}}FB_{22}^{\beta,a,b}F\\
\sim e^{D_{\text{even}}}\left(\frac{1}{(2n+a+b+1)^{-1}}FB_{22}^{\infty,a,b}F\right)e^{-D_{\text{odd}}}\label{eq:identityforjacobihard}\end{multline}
with\begin{align*}
D_{\text{even}} & =\diag(d_{2},d_{4},\dots,d_{2n}),\; D_{\text{odd}}=\diag(d_{1},d_{3},\dots,d_{2n-1})\\
d_{i} & ={\textstyle \sum_{k=i}^{2n-1}g_{k}}\\
g_{1} & \sim-(\log c_{1}-\log\bar{c}_{1})\\
g_{2i} & \sim(\log s_{i}-\log\bar{s}_{i})+(\log c'_{i}-\log\bar{c}'_{i})\\
g_{2i-1} & \sim-(\log c_{i}-\log\bar{c}_{i})-(\log s'_{i-1}-\log\bar{s}'_{i-1})\;\text{for }i\geq2\\
\cos^{2}\theta_{i} & \sim\betapdf({\textstyle \frac{\beta}{2}(a+i),\frac{\beta}{2}(b+i)})\\
\cos^{2}\phi_{i} & \sim\betapdf({\textstyle \frac{\beta}{2}i,\frac{\beta}{2}(a+b+1+i)}).\end{align*}
 The random angles $\theta_{1},\dots,\theta_{n}$ and $\phi_{1},\dots,\phi_{n-1}$
are independent, and their cosines and sines are denoted by $c_{i}$,
$s_{i}$, $c'_{i}$, and $s'_{i}$, as usual. The constants $\bar{c}_{i}$,
$\bar{s}_{i}$, $\bar{c}'_{i}$, and $\bar{s}'_{i}$ are introduced
after the definition of the $\beta$-Jacobi matrix model in Section
\ref{sec:jacobimodel}.

\section{\label{sec:zerotempasfinitediff}Zero temperature matrix models as
finite difference schemes}

As seen in Section \ref{sec:matrixmodels}, the Hermite, Laguerre,
and Jacobi matrix models approach nonrandom limits as $\beta\rightarrow\infty$.
We call these matrices {}``zero temperature matrix models'' because
of the well known connection with statistical mechanics.

By appropriately transforming the zero temperature matrix models---via
operations such as translation, scalar multiplication, similarity
transform, and factorization---we can interpret them as finite difference
schemes for the classical Airy and Bessel operators. This approach
anticipates analogous methods for the $\beta<\infty$ case. In short,
$\beta=\infty$ matrix models discretize nonrandom operators, and
$\beta<\infty$ matrix models discretize stochastic operators.

\subsection{Soft edge}

\subsubsection{Hermite $\rightarrow$ Airy}

\begin{defn}
Let $h=n^{-1/3}$. The $n$-by-$n$ \emph{$\infty$-Hermite matrix
model scaled at the soft edge} is\[
H_{\text{soft}}^{\infty}=-\sqrt{\frac{2}{h}}(DH^{\infty}D^{-1}-\sqrt{2}h^{-3/2}I),\]
in which $DH^{\infty}D^{-1}$ is the matrix of (\ref{eq:infhermiteaftersimilarity}).
\end{defn}
Note that {}``scaling at the soft edge'' modifies eigenvalues in
a benign way. The translation and rescaling are designed so that the
smallest $k$ eigenvalues of $H_{\text{soft}}^{\infty}$ approach
distinct limits as $n\rightarrow\infty$. (The largest eigenvalues
of $DH^{\infty}D^{-1}$ are first pulled toward the origin, and then
a scalar factor is applied to {}``zoom in.'' The scalar factor is
negative to produce an increasing, as opposed to decreasing, sequence
of eigenvalues starting near zero.)

The following theorem interprets $H_{\text{soft}}^{\infty}$ as a
finite difference scheme for the classical Airy operator $\mathcal{A}^{\infty}=-\frac{d^{2}}{dx^{2}}+x$
on the mesh $x_{i}=hi$, $i=1,\dots,n$, with mesh size $h=n^{-1/3}$.

\begin{thm}
\label{thm:infhermitetoairy}For all positive integers $n$,\[
H_{\text{soft}}^{\infty}=\frac{1}{h^{2}}\Delta+\diag_{-1}(x_{1},x_{2},\dots,x_{n-1}),\]
in which $h=n^{-1/3}$, $x_{i}=hi$, and $\diag_{-1}(x_{1},x_{2},\dots,x_{n-1})$
is the $n$-by-$n$ matrix with $x_{1},x_{2},\dots,x_{n-1}$ on the
subdiagonal and zeros elsewhere.

Furthermore, for fixed $k$, the $k$th least eigenvalue of $H_{\text{soft}}^{\infty}$
converges to the $k$th least eigenvalue of $\mathcal{A}^{\infty}$
as $n\rightarrow\infty$,\[
\lambda_{k}(H_{\text{soft}}^{\infty})\stackrel{n\rightarrow\infty}{\rightarrow}\lambda_{k}(\mathcal{A}^{\infty}).\]

\end{thm}
\begin{proof}
The expression for $H_{\text{soft}}^{\infty}$ is straightforward
to derive. For the eigenvalue result, recall that the $k$th greatest
eigenvalue of $H^{\infty}$ is the $k$th rightmost root of the $n$th
Hermite polynomial, and the $k$th least eigenvalue of $\mathcal{A}^{\infty}$
is the $k$th zero of $\Ai$, up to sign. The eigenvalue convergence
result is exactly equation (6.32.5) of \cite{MR0372517}. (The recentering
and rescaling in the definition of $H_{\text{soft}}^{\infty}$ is
designed precisely for the purpose of applying that equation.) 
\end{proof}
It is also true that the eigenvectors of $H_{\text{soft}}^{\infty}$
discretize the eigenfunctions of $\mathcal{A}^{\infty}$. This can
be established with well known orthogonal polynomial asymptotics,
specifically equation (3.3.23) of \cite{MR0372517}. We omit a formal
statement and proof for brevity's sake.

\subsubsection{Laguerre $\rightarrow$ Airy}

\begin{defn}
Let $h=\left(2n\right)^{-1/3}$. The $2n$-by-$2n$ \emph{$\infty$-Laguerre
matrix model scaled at the soft edge} is\begin{align*}
L_{\text{soft}}^{\infty,a} & =-\sqrt{\frac{2}{h}}\left(D_{L}P_{L}^{T}\left[\begin{array}{cc}
0 & (L^{\infty,a})^{T}\\
L^{\infty,a} & 0\end{array}\right]P_{L}D_{L}^{-1}-\sqrt{2}h^{-3/2}I\right),\end{align*}
in which $D_{L}$ is the matrix $D$ of Lemma \ref{lem:identityforlairy}
(with $\beta=\infty$) and $P_{L}$ is the matrix $P$ of the same
lemma. The $(2n+1)$-by-$(2n+1)$ \emph{$\infty$-Laguerre matrix
model scaled at the soft edge} is\begin{align*}
M_{\text{soft}}^{\infty,a} & =-\sqrt{\frac{2}{h}}\left(D_{M}P_{M}^{T}\left[\begin{array}{cc}
0 & M^{\infty,a}\\
(M^{\infty,a})^{T} & 0\end{array}\right]P_{M}D_{M}^{-1}-\sqrt{2}h^{-3/2}I\right),\end{align*}
in which $D_{M}$ is the matrix $D$ of Lemma \ref{lem:identityformairy}
(with $\beta=\infty$) and $P_{M}$ is the matrix $P$ of the same
lemma.
\end{defn}
$L_{\text{soft}}^{\infty,a}$ and $M_{\text{soft}}^{\infty,a}$ encode
finite difference schemes for the classical Airy operator $\mathcal{A}^{\infty}=-\frac{d^{2}}{dx^{2}}+x$.

\begin{thm}
\label{thm:inflaguerretoairy}Make the approximations\begin{align*}
L_{\text{soft}}^{\infty,a} & =\frac{1}{h^{2}}\Delta+\diag_{-1}(x_{1},x_{2},\dots,x_{2n-1})+E_{L}\\
\intertext{and}M_{\text{soft}}^{\infty,a} & =\frac{1}{h^{2}}\Delta+\diag_{-1}(x_{0},x_{1},x_{2},\dots,x_{2n-1})+E_{M},\end{align*}
with $h=(2n)^{-1/3}$, $x_{i}=hi$, and $\diag_{-1}(d_{1},\dots,d_{m})$
denoting the square matrix with $d_{1},\dots,d_{m}$ on the subdiagonal
and zeros elsewhere. Then $E_{L}$ and $E_{M}$ are zero away from
the subdiagonal, and their entries are uniformly $O(h)$ as $n\rightarrow\infty$.
In the special case $a=-\frac{1}{2}$, $E_{L}$ equals the zero matrix,
and in the special case $a=\frac{1}{2}$, $E_{M}$ equals the zero
matrix.

Furthermore, for fixed $k$, the $k$th least eigenvalue of $L_{\text{soft}}^{\infty,a}$
and the $k$th least eigenvalue of $M_{\text{soft}}^{\infty,a}$ converge
to the $k$th least eigenvalue of the classical operator as $n\rightarrow\infty$,\begin{align*}
\lambda_{k}(L_{\text{soft}}^{\infty,a}) & \stackrel{n\rightarrow\infty}{\rightarrow}\lambda_{k}(\mathcal{A}^{\infty}),\\
\lambda_{k}(M_{\text{soft}}^{\infty,a}) & \stackrel{n\rightarrow\infty}{\rightarrow}\lambda_{k}(\mathcal{A}^{\infty}).\end{align*}

\end{thm}
\begin{proof}
For odd $j$, the $(j+1,j)$ entry of $E_{L}$ equals $-h(2a+1)$,
and every other entry of $E_{L}$ equals zero. For even $j$, the
$(j+1,j)$ entry of $E_{M}$ equals $-h(2a-1)$, and every other entry
of $E_{M}$ equals zero. For the eigenvalue result, check that the
$k$th greatest eigenvalue of (\ref{eq:shuffledl}), resp., (\ref{eq:shuffledm}),
equals the $k$th greatest singular value of $L^{\infty,a}$, resp.,
$M^{\infty,a}$, and that this value is the square root of the $k$th
rightmost root of the $n$th Laguerre polynomial with parameter $a$.
The eigenvalue convergence then follows from equation (6.32.4) of
\cite{MR0372517}, concerning zero asymptotics for Laguerre polynomials.
\end{proof}
Also, the $k$th eigenvector of $L_{\text{soft}}^{\infty,a}$, resp.,
$M_{\text{soft}}^{\infty,a}$, discretizes the $k$th eigenfunction
of $\mathcal{A}^{\infty}$, via (3.3.21) of \cite{MR0372517}.

\subsection{Hard edge}

\subsubsection{Laguerre $\rightarrow$ Bessel}

\begin{defn}
The $n$-by-$n$ \emph{$\infty$-Laguerre matrix model scaled at the
hard edge} is\[
L_{\text{hard}}^{\infty,a}=\sqrt{\frac{2}{h}}F\Omega(L^{\infty,a})^{T}\Omega F,\]
in which $h=\frac{1}{2n+a+1}$. $F$ and $\Omega$ are defined in
Section \ref{section:backgroundfinitediffs}.
\end{defn}
Note that {}``scaling at the hard edge'' only modifies singular
values by a constant factor. This factor is chosen so that the $k$
smallest singular values approach distinct limits as $n\rightarrow\infty$.

The next theorem interprets $L_{\text{hard}}^{\infty,a}$ as a discretization
of the classical Bessel operator $\mathcal{J}_{a}^{\infty}=-2\sqrt{x}\frac{d}{dx}+\frac{a}{\sqrt{x}}$
with type (i) boundary conditions. Domain and codomain vectors are
interpreted on interwoven submeshes. The combined mesh has size $h=\frac{1}{2n+a+1}$
and grid points $x_{i}=h(a+i)$, $i=1,\dots,2n$. Domain vectors are
interpreted on the submesh $x_{2i-1}$, $i=1,\dots,n$, and codomain
vectors are interpreted on the submesh $x_{2i}$, $i=1,\dots,n$. 

\begin{thm}
\label{thm:inflaguerreltobessel}Let $h$ and $x_{i}$ be defined
as in the previous paragraph, and make the approximation\begin{multline*}
L_{\text{hard}}^{\infty,a}=-2\diag(\sqrt{x_{2}},\sqrt{x_{4}},\dots,\sqrt{x_{2n}})({\textstyle \frac{1}{2h}}\nabla)\\
+a\diag\left(\frac{1}{\sqrt{x_{2}}},\frac{1}{\sqrt{x_{4}}},\dots,\frac{1}{\sqrt{x_{2n}}}\right)S+E,\end{multline*}
with $S$ defined as in Section \ref{section:backgroundfinitediffs}.
Then the error term $E$ is upper bidiagonal, and the entries in rows
$\lceil\frac{\varepsilon}{h}\rceil,\dots,n$ of $E$ are uniformly
$O(h)$, for any fixed $\varepsilon>0$.

Furthermore, for fixed $k$, the $k$th least singular value of $L_{\text{hard}}^{\infty,a}$
approaches the $k$th least singular value of $\mathcal{J}_{a}^{\infty}$
with type (i) boundary conditions as $n\rightarrow\infty$,\[
\sigma_{k}(L_{\text{hard}}^{\infty,a})\stackrel{n\rightarrow\infty}{\rightarrow}\sigma_{k}(\mathcal{J}_{a}^{\infty}\text{ with type (i) b.c.'s}).\]

\end{thm}
\begin{proof}
The $(i,i)$ entry of $E$ equals $\frac{1}{h}\sqrt{x_{2i}+ha}-\frac{1}{h}\sqrt{x_{2i}}-\frac{a}{2}x_{2i}^{-1/2}$.
By a Taylor series expansion, $\sqrt{x+ha}=\sqrt{x}+\frac{ha}{2}x^{-1/2}+O(h^{2})$,
uniformly for any set of $x$ values bounded away from zero. This
implies that the $(i,i)$ entry of $E$ is $O(h)$, for any sequence
of values for $i$ bounded below by $\lceil\frac{\varepsilon}{h}\rceil$
as $n\rightarrow\infty$. The $(i,i+1)$ entry of $E$ equals $-\frac{1}{h}\sqrt{x_{2i}-ha}+\frac{1}{h}\sqrt{x_{2i}}-\frac{a}{2}x_{2i}^{-1/2}$,
from which similar asymptotics follow. For the singular value result,
recall that the $k$th least singular value of $L^{\infty,a}$ is
the square root of the $k$th least root of the Laguerre polynomial
with parameter $a$ and that the $k$th least singular value of $\mathcal{J}_{a}^{\infty}$
is the $k$th positive zero of $j_{a}$, the Bessel function of the
first kind of order $a$. The convergence result follows immediately
from (6.31.6) of \cite{MR0372517}.
\end{proof}
In fact, the singular vectors of $L_{\text{hard}}^{\infty,a}$ discretize
the singular functions of $\mathcal{J}_{a}^{\infty}$ with type (i)
boundary conditions as well. This can be proved with (3.3.20) of \cite{MR0372517}.

Analogous results hold for the rectangular $\beta$-Laguerre matrix
model.

\begin{defn}
The $n$-by-$(n+1)$ \emph{$\infty$-Laguerre matrix model scaled
at the hard edge} is\[
M_{\text{hard}}^{\infty,a}=-\sqrt{\frac{2}{h}}F_{n}\Omega_{n}(M^{\infty,a})^{T}\Omega_{n+1}F_{n+1},\]
 in which $h=\frac{1}{2n+a+1}$. $F$ and $\Omega$ are defined in
Section \ref{section:backgroundfinitediffs}.
\end{defn}
The next theorem interprets $M_{\text{hard}}^{\infty,a}$ as a finite
difference scheme for the classical Bessel operator $\mathcal{J}_{a-1}^{\infty}=-2\sqrt{x}\frac{d}{dx}+\frac{a-1}{\sqrt{x}}$,
with type (ii) boundary conditions. Domain and codomain vectors are
interpreted on interwoven submeshes. The combined mesh has size $h=\frac{1}{2n+a+1}$
and grid points $x_{i}=h(a-1+i)$, $i=1,\dots,2n+1$. Domain vectors
are interpreted on the submesh $x_{2i-1}$, $i=1,\dots,n+1$, and
codomain vectors are interpreted on the submesh $x_{2i}$, $i=1,\dots,n$.

\begin{thm}
\label{thm:inflaguerremtobessel} Let $h$ and $x_{i}$ be defined
as in the previous paragraph, and make the approximation\begin{multline*}
M_{\text{hard}}^{\infty,a}=-2\diag(\sqrt{x_{2}},\sqrt{x_{4}},\dots,\sqrt{x_{2n}})({\textstyle \frac{1}{2h}}\nabla_{n,n+1})\\
+(a-1)\diag\left(\frac{1}{\sqrt{x_{2}}},\frac{1}{\sqrt{x_{4}}},\dots,\frac{1}{\sqrt{x_{2n}}}\right)S_{n,n+1}+E,\end{multline*}
with $S$ defined as in Section \ref{section:backgroundfinitediffs}.
Then the error term $E$ is upper bidiagonal, and the entries in rows
$\lceil\frac{\varepsilon}{h}\rceil,\dots,n$ of $E$ are uniformly
$O(h)$, for any fixed $\varepsilon>0$.

Furthermore, for fixed $k$, the $k$th least singular value of $M_{\text{hard}}^{\infty,a}$
approaches the $k$th least singular value of $\mathcal{J}_{a-1}^{\infty}$
with type (ii) boundary conditions,\[
\sigma_{k}(M_{\text{hard}}^{\infty,a})\stackrel{n\rightarrow\infty}{\rightarrow}\sigma_{k}(\mathcal{J}_{a-1}^{\infty}\text{ with type (ii) b.c.'s}).\]

\end{thm}
\begin{proof}
The $(i,i)$ entry of $E$ equals $\frac{1}{h}\sqrt{x_{2i}+h(a-1)}-\frac{1}{h}\sqrt{x_{2i}}-\frac{(a-1)}{2}x_{2i}^{-1/2}$.
By a Taylor series expansion, $\sqrt{x+h(a-1)}=\sqrt{x}+\frac{h(a-1)}{2}x^{-1/2}+O(h^{2})$,
uniformly for any set of $x$ values bounded away from zero. This
implies that the $(i,i)$ entry of $E$ is $O(h)$, for any sequence
of values for $i$ bounded below by $\lceil\frac{\varepsilon}{h}\rceil$
as $n\rightarrow\infty$. The $(i,i+1)$ entry of $E$ equals $-\frac{1}{h}\sqrt{x_{2i}-h(a-1)}+\frac{1}{h}\sqrt{x_{2i}}-\frac{a-1}{2}x_{2i}^{-1/2}$,
from which similar asymptotics follow. For the singular value result,
the proof of Theorem \ref{thm:inflaguerreltobessel} suffices, because
$L_{\text{hard}}^{\infty,a}$ and $M_{\text{hard}}^{\infty,a}$ have
exactly the same nonzero singular values, as do $\mathcal{J}_{a}^{\infty}$
with type (i) boundary conditions and $\mathcal{J}_{a-1}^{\infty}$
with type (ii) boundary conditions.
\end{proof}
Also, the singular vectors of $M_{\text{hard}}^{\infty,a}$ discretize
the singular functions of $\mathcal{J}_{a-1}^{\infty}$ with type
(ii) b.c.'s, although we omit a formal statement of this fact here.

\subsubsection{Jacobi $\rightarrow$ Bessel}

\begin{defn}
The $n$-by-$n$ \emph{$\infty$-Jacobi matrix model scaled at the
hard edge} is\[
J_{\text{hard}}^{\infty,a,b}=\frac{1}{h}FB_{22}^{\infty,a,b}F,\]
in which $h=\frac{1}{2n+a+b+1}$ and $B_{22}^{\infty,a,b}$ is the
bottom-right block of the $2n$-by-$2n$ $\infty$-Jacobi matrix model
$J^{\infty,a,b}$. $F$ and $\Omega$ are defined in Section \ref{section:backgroundfinitediffs}.
\end{defn}
The next theorem interprets $J_{\text{hard}}^{\infty,a,b}$ as a discretization
of the classical Bessel operator in Liouville normal form, $\tilde{\mathcal{J}}_{a}^{\infty}=-\frac{d}{dx}+(a+\frac{1}{2})\frac{1}{x}$
with type (i) boundary conditions. Domain and codomain vectors are
interpreted on interwoven meshes. The combined mesh has size $h=\frac{1}{2n+a+b+1}$
and grid points $x_{i}=h(a+b+i)$, $i=1,\dots,2n$. Domain vectors
are interpreted on the mesh $x_{2i-1}$, $i=1,\dots,n$, and codomain
vectors are interpreted on the mesh $x_{2i}$, $i=1,\dots,n$.

\begin{thm}
\label{thm:infjacobitobessel}Let $h$ and $x_{i}$ be defined as
in the previous paragraph, and make the approximation\[
J_{\text{hard}}^{\infty,a,b}=-\frac{1}{2h}\nabla+\left(a+\frac{1}{2}\right)\diag\left(\frac{1}{x_{2}},\frac{1}{x_{4}},\dots,\frac{1}{x_{2n}}\right)S+E,\]
with $S$ defined as in Section \ref{section:backgroundfinitediffs}.
Then the error term $E$ is upper bidiagonal, and the entries in rows
$\lceil\frac{\varepsilon}{h}\rceil,\dots,n$ are uniformly $O(h)$,
for any fixed $\varepsilon>0$.

Furthermore, for fixed $k$, the $k$th least singular value of $J_{\text{hard}}^{\infty,a,b}$
approaches the $k$th least singular value of $\tilde{\mathcal{J}}_{a}^{\infty}$
with type (i) boundary conditions as $n\rightarrow\infty$,\[
\sigma_{k}(J_{\text{hard}}^{\infty,a,b})\stackrel{n\rightarrow\infty}{\rightarrow}\sigma_{k}(\tilde{\mathcal{J}}_{a}^{\infty}\text{ with type (i) b.c.'s}).\]

\end{thm}
\begin{proof}
The $(i,i)$ entry of $J_{\text{hard}}^{\infty,a,b}$ equals \begin{multline*}
\frac{1}{h}\sqrt{\frac{a+i}{a+b+2i}}\sqrt{\frac{a+b+i}{a+b+1+2(i-1)}}\\
=\frac{1}{2h}(x_{2i}+h(a-b))^{1/2}x_{2i}^{-1/2}(x_{2i}+h(a+b))^{1/2}(x_{2i}-h)^{-1/2}.\end{multline*}
Rewriting this expression as\[
\frac{1}{2h}\left((x_{2i}+ha)^{2}-h^{2}b^{2}\right)^{1/2}\left((x_{2i}-{\textstyle \frac{h}{2}})^{2}-{\textstyle \frac{h^{2}}{4}}\right)^{-1/2},\]
 it is straightforward to check that the entry is $\frac{1}{2h}+(a+{\textstyle \frac{1}{2}})\cdot\frac{1}{2}\frac{1}{x_{2i}}+O(h)$,
uniformly for any sequence of values $i$ such that $x_{2i}$ is bounded
away from zero as $n\rightarrow\infty$. The argument for the superdiagonal
terms is similar.

For the singular value result, note that the CS values of $J^{\infty,a,b}$
equal the singular values of its bottom-right block, and that these
values, squared, equal the roots of the $n$th Jacobi polynomial with
parameters $a$, $b$. Also recall that the $k$th least singular
value of $\tilde{\mathcal{J}}_{a}^{\infty}$ with type (i) boundary
conditions is the $k$th positive zero of $j_{a}$, the Bessel function
of the first kind of order $a$. The rescaling in the definition of
$J_{\text{hard}}^{\infty,a,b}$ is designed so that equation (6.3.15)
of \cite{MR0372517} may be applied at this point, proving convergence.
\end{proof}
It is also true that the singular vectors of $J_{\text{hard}}^{\infty,a,b}$
discretize the singular functions of $\tilde{\mathcal{J}}_{a}^{\infty}$
with type (i) boundary conditions.

As presented here, the theorem only considers the bottom-right block
of $J^{\infty,a,b}$, but similar estimates have been derived for
the other three blocks \cite{mythesis}. Briefly, the bottom-right
and top-left blocks discretize $\tilde{\mathcal{J}}_{a}^{\infty}$
and $(\tilde{\mathcal{J}}_{a}^{\infty})^{*}$, respectively, while
the top-right and bottom-left blocks discretize $\tilde{\mathcal{J}}_{b}^{\infty}$
and $(\tilde{\mathcal{J}}_{b}^{\infty})^{*}$, respectively, all with
type (i) boundary conditions.

\section{\label{sec:randomasfinitediff}Random matrix models as finite difference
schemes}

The previous section demonstrated how to view zero temperature matrix
models as finite difference schemes for differential operators. Because
the matrices were not random, the differential operators were not
random either. This section extends to the finite $\beta$ case, when
randomness appears.

\subsection{Soft edge}

\subsubsection{Hermite $\rightarrow$ Airy}

\begin{defn}
Let $h=n^{-1/3}$. The $n$-by-$n$ \emph{$\beta$-Hermite matrix
model scaled at the soft edge} is\[
H_{\text{soft}}^{\beta}\sim-\sqrt{\frac{2}{h}}(DH^{\beta}D^{-1}-\sqrt{2}h^{-3/2}I),\]
with $DH^{\beta}D^{-1}$ denoting the matrix in (\ref{eq:hermiteaftersimilarity}).
\end{defn}
The eigenvalues of $H_{\text{soft}}^{\beta}$ display soft edge behavior
as $n\rightarrow\infty$. The underlying reason, we claim, is that
the matrix is a discretization of the stochastic Airy operator. The
next theorem interprets $H_{\text{soft}}^{\beta}$ as a finite difference
scheme with mesh size $h=n^{-1/3}$ and grid points $x_{i}=hi$, $i=1,\dots,n$.

\begin{thm}
\label{thm:betahermitetoairy}We have\[
H_{\text{soft}}^{\beta}\sim H_{\text{soft}}^{\infty}+\frac{2}{\sqrt{\beta}}W,\]
in which\[
W\sim\frac{1}{\sqrt{h}}\left(\frac{-1}{\sqrt{2}}\right)\left[\begin{array}{ccccc}
G_{1}\\
\tilde{\chi}_{(n-1)\beta}^{2} & G_{2}\\
 & \tilde{\chi}_{(n-2)\beta}^{2} & G_{3}\\
 &  & \ddots & \ddots\\
 &  &  & \tilde{\chi}_{\beta}^{2} & G_{n}\end{array}\right].\]
Here, $h=n^{-1/3}$, $W$ has independent entries, $G_{1},\dots,G_{n}$
are standard Gaussian random variables, and $\tilde{\chi}_{(n-j)\beta}^{2}\sim\frac{1}{\sqrt{2\beta n}}(\chi_{(n-j)\beta}^{2}-(n-j)\beta)$.
Further, $\tilde{\chi}_{(n-j)\beta}^{2}$ has mean zero and standard
deviation $1+O(h^{2})$, uniformly for $j$ such that $x_{j}=hj\leq M$,
where $M>0$ is fixed.
\end{thm}
\begin{proof}
The derivation of the expression for $H_{\text{soft}}^{\beta}$ is
straightforward. The mean of $\tilde{\chi}_{(n-j)\beta}^{2}$ is exactly
0, and the variance is exactly $1-h^{2}x_{j}$.
\end{proof}
We claim that the matrix $W$ discretizes white noise on the mesh
from Theorem \ref{thm:infhermitetoairy}. The increment of Brownian
motion over an interval $(x,x+h]$ has mean 0 and standard deviation
$\sqrt{h}$, so a discretization of white noise over the same interval
should have mean 0 and standard devation $\frac{1}{\sqrt{h}}$. The
noise in the matrix $W$ has the appropriate mean and standard deviation.

\subsubsection{Laguerre $\rightarrow$ Airy}

\begin{defn}
\label{def:betalaguerreatsoft}Let $h=\left(2n\right)^{-1/3}$. The
$2n$-by-$2n$ \emph{$\beta$-Laguerre matrix model scaled at the
soft edge} is\begin{align*}
L_{\text{soft}}^{\beta,a} & \sim-\sqrt{\frac{2}{h}}\left(D_{L}P_{L}^{T}\left[\begin{array}{cc}
0 & (L^{\beta,a})^{T}\\
L^{\beta,a} & 0\end{array}\right]P_{L}D_{L}^{-1}-\sqrt{2}h^{-3/2}I\right),\end{align*}
in which $D_{L}$ is the matrix $D$ of Lemma \ref{lem:identityforlairy}
and $P_{L}$ is the matrix $P$ of the same lemma. The $(2n+1)$-by-$(2n+1)$
\emph{$\beta$-Laguerre matrix model scaled at the soft edge} is\begin{align*}
M_{\text{soft}}^{\beta,a} & \sim-\sqrt{\frac{2}{h}}\left(D_{M}P_{M}^{T}\left[\begin{array}{cc}
0 & M^{\beta,a}\\
(M^{\beta,a})^{T} & 0\end{array}\right]P_{M}D_{M}^{-1}-\sqrt{2}h^{-3/2}I\right),\end{align*}
in which $D_{M}$ is the matrix $D$ of Lemma \ref{lem:identityformairy}
and $P_{M}$ is the matrix $P$ of the same lemma.
\end{defn}
The eigenvalues of $L_{\text{soft}}^{\beta,a}$ and $M_{\text{soft}}^{\beta,a}$
near zero display soft edge behavior as $n\rightarrow\infty$. The
underlying reason, we claim, is that the matrices themselves encode
finite difference schemes for the stochastic Airy operator, as the
next theorem shows.

\begin{thm}
\label{thm:betalaguerretoairy}The $2n$-by-$2n$ and $(2n+1)$-by-$(2n+1)$
$\beta$-Laguerre matrix models scaled at the soft edge satisfy\begin{align*}
L_{\text{soft}}^{\beta,a} & \sim L_{\text{soft}}^{\infty,a}+\frac{2}{\sqrt{\beta}}W_{L}\\
\intertext{and}M_{\text{soft}}^{\beta,a} & \sim M_{\text{soft}}^{\infty,a}+\frac{2}{\sqrt{\beta}}W_{M},\end{align*}
with\begin{align*}
W_{L} & \sim\frac{-1}{\sqrt{h}}\diag_{-1}(\tilde{\chi}_{(a+n)\beta}^{2},\tilde{\chi}_{(n-1)\beta}^{2},\tilde{\chi}_{(a+n-1)\beta}^{2},\tilde{\chi}_{(n-2)\beta}^{2},\dots\\
 & \qquad\qquad\qquad\qquad\qquad\qquad\qquad\qquad\qquad\quad\dots,\tilde{\chi}_{(a+2)\beta}^{2},\tilde{\chi}_{\beta}^{2},\tilde{\chi}_{(a+1)\beta}^{2}),\\
W_{M} & \sim\frac{-1}{\sqrt{h}}\diag_{-1}(\tilde{\chi}_{n\beta}^{2},\tilde{\chi}_{(a+n-1)\beta}^{2},\tilde{\chi}_{(n-1)\beta}^{2},\tilde{\chi}_{(a+n-2)\beta}^{2},\dots\\
 & \qquad\qquad\qquad\qquad\qquad\qquad\qquad\qquad\qquad\quad\dots,\tilde{\chi}_{2\beta}^{2},\tilde{\chi}_{(a+1)\beta}^{2},\tilde{\chi}_{\beta}^{2},\tilde{\chi}_{a\beta}^{2}),\end{align*}
and $h=(2n)^{-1/3}$. All $2n-1$ subdiagonal entries of $W_{L}$
and all $2n$ subdiagonal entries of $W_{M}$ are independent, with
$\tilde{\chi}_{r}^{2}$ denoting a random variable with distribution
$\tilde{\chi}_{r}^{2}\sim\frac{1}{\sqrt{2\beta n}}(\chi_{r}^{2}-r)$.
\end{thm}
The entries of $W_{L}$ and $W_{M}$ have mean approximately 0 and
standard deviation approximately $\frac{1}{\sqrt{h}}$. Therefore,
we think of $W_{L}$ and $W_{M}$ as discretizations of white noise
on the mesh from Theorem \ref{thm:inflaguerretoairy}. The situation
is very similar to that in Theorem \ref{thm:betahermitetoairy}, so
we omit a formal statement.

\subsubsection{Overview of finite difference schemes for the stochastic Airy operator}

In light of Theorems \ref{thm:infhermitetoairy}, \ref{thm:inflaguerretoairy},
\ref{thm:betahermitetoairy}, and \ref{thm:betalaguerretoairy}, we
make the following claim. $H_{\text{soft}}^{\beta}$, $L_{\text{soft}}^{\beta,a}$,
and $M_{\text{soft}}^{\beta,a}$ discretize the stochastic Airy operator
$\mathcal{A}^{\beta}=-\frac{d^{2}}{dx^{2}}+x+\frac{2}{\sqrt{\beta}}B'$,
for finite and infinite $\beta$. Because the low eigenvalues of $H_{\text{soft}}^{\beta}$,
$L_{\text{soft}}^{\beta,a}$, and $M_{\text{soft}}^{\beta,a}$ show
soft edge behavior, it is natural to expect that the eigenvalues of
$\mathcal{A}^{\beta}$ show the same behavior.

\begin{conjecture}
\label{conj:airyeigenvalues} The $k$th least eigenvalue of the stochastic
Airy operator follows the $k$th soft edge distribution, with the
same value for $\beta$.
\end{conjecture}
The conjecture now appears to be a theorem, due to a proof of Ram\'{i}rez,
Rider, and Vir\'{a}g \cite{rideretal}.

\subsection{Hard edge}

\subsubsection{Laguerre $\rightarrow$ Bessel}

\begin{defn}
The $n$-by-$n$ \emph{$\beta$-Laguerre matrix model scaled at the
hard edge} is\[
L_{\text{hard}}^{\beta,a}\sim\sqrt{\frac{2}{h}}F\Omega(L^{\beta,a})^{T}\Omega F,\]
in which $h=\frac{1}{2n+a+1}$ and $F$ and $\Omega$ are defined
as in Section \ref{section:backgroundfinitediffs}.
\end{defn}
The least singular values of $L_{\text{hard}}^{\beta,a}$ display
hard edge behavior as $n\rightarrow\infty$. We claim that this can
be understood by viewing the matrix as a finite difference scheme
for the stochastic Bessel operator with type (i) boundary conditions.
The next theorem demonstrates this, using the same mesh seen in Theorem
\ref{thm:inflaguerreltobessel}.

{}

\begin{thm}
\label{thm:betalaguerretobessel}Let $L_{\text{hard}}^{\beta,a}$
be a matrix from the $n$-by-$n$ $\beta$-Laguerre matrix model scaled
at the hard edge. Adopting the notation of (\ref{eq:identityforlhard})
and Theorem \ref{thm:inflaguerreltobessel} and setting $\tilde{g}_{i}=-\sqrt{\beta x_{i}/h}\, g_{i}$,
we have
\begin{enumerate}
\item $L_{\text{hard}}^{\beta,a}\sim e^{D_{\text{even}}}L_{\text{hard}}^{\infty,a}e^{-D_{\text{odd}}}.$
\item $e^{d_{i}}=\exp\left(-\frac{1}{\sqrt{\beta}}\sum_{k=i}^{2n-1}x_{k}^{-1/2}(\tilde{g}_{k}\sqrt{h})\right)$.
\item $\tilde{g}_{1},\dots,\tilde{g}_{2n-1}$ are independent, and, for
any $\varepsilon>0$, the random variables $\tilde{g}_{\lceil\varepsilon/h\rceil},\dots,\tilde{g}_{2n-1}$
have mean $O(\sqrt{h})$ and standard deviation $1+O(h)$, uniformly.
\end{enumerate}
\end{thm}
The point of (3) is that the sequence $\frac{1}{\sqrt{h}}\tilde{g}_{1},\dots,\frac{1}{\sqrt{h}}\tilde{g}_{2n-1}$
is a discretization of white noise. Hence, the expression for $e^{d_{i}}$
in (2) is a discretization of $\psi(x_{i})$.

\begin{proof}
Conclusions (1) and (2) are simply restatements of facts from (\ref{eq:identityforlhard}).
For (3), the independence of $\tilde{g}_{1},\dots,\tilde{g}_{2n-1}$
was already established in the context of (\ref{eq:identityforlhard}).
For the asymptotic mean and standard deviation, use the fact that
$\log(\chi_{r}/\sqrt{r})$ has mean $-\frac{1}{2}\log(r/2)+\frac{\Gamma'(r/2)}{\Gamma(r/2)}=O(r^{-1})$
and standard deviation $\frac{1}{\sqrt{2r}}+O(r^{-3/2})$ as $r\rightarrow\infty$.
\end{proof}
Theorem \ref{thm:inflaguerreltobessel} establishes that $L_{\text{hard}}^{\infty,a}$
can be viewed as a finite difference scheme for the classical Bessel
operator,\[
L_{\text{hard}}^{\infty,a}\;\stackrel{n\rightarrow\infty}{\rightsquigarrow}\;\mathcal{J}_{a}^{\infty}.\]
 Theorem \ref{thm:betalaguerretobessel} extends this connection to
the $\beta<\infty$ case. Noting the apparent convergence in distribution
\[
\exp\left(-\frac{1}{\sqrt{\beta}}\sum_{k=\lceil x-x_{0}\rceil/h}^{2n-1}x_{k}^{-1/2}(\tilde{g}_{k}\sqrt{h})\right)\;\stackrel{n\rightarrow\infty}{\rightsquigarrow}\;\psi(x),\]
 $L_{\text{hard}}^{\beta,a}$ appears to be a finite difference scheme
for the stochastic Bessel operator with type (i) boundary conditions,\[
L_{\text{hard}}^{\beta,a}\;\sim\; e^{D_{\text{even}}}L_{\text{hard}}^{\infty,a}e^{-D_{\text{odd}}}\;\stackrel{n\rightarrow\infty}{\rightsquigarrow}\;\psi\mathcal{J}_{a}^{\infty}\psi^{-1}\;\sim\;\mathcal{J}_{a}^{\beta}.\]

\begin{defn}
The $n$-by-$(n+1)$ \emph{$\beta$-Laguerre matrix model scaled at
the hard edge} is\[
M_{\text{hard}}^{\beta,a}\sim-\sqrt{\frac{2}{h}}F_{n}\Omega_{n}(M^{\beta,a})^{T}\Omega_{n+1}F_{n+1},\]
in which $h=\frac{1}{2n+a+1}$. $F$ and $\Omega$ are defined in
Section \ref{section:backgroundfinitediffs}.
\end{defn}
The small singular values of $M_{\text{hard}}^{\beta,a}$ display
hard edge behavior as $n\rightarrow\infty$, because, we claim, the
matrix is a finite difference scheme for the stochastic Bessel operator
with type (ii) boundary conditions. The next theorem demonstrates
this, using the same mesh seen in Theorem \ref{thm:inflaguerremtobessel}.

\begin{thm}
\label{thm:betalaguerremtobessel}Let $M_{\text{hard}}^{\beta,a}$
be a matrix from the $n$-by-$(n+1)$ $\beta$-Laguerre matrix model
scaled at the hard edge. Adopting the notation of (\ref{eq:identityformhard})
and Theorem \ref{thm:inflaguerremtobessel} and setting $\tilde{g}_{i}=-\sqrt{\beta x_{i}/h}\, g_{i}$,
we have
\begin{enumerate}
\item $M_{\text{hard}}^{\beta,a}\sim e^{D_{\text{even}}}M_{\text{hard}}^{\infty,a}e^{-D_{\text{odd}}}.$
\item $e^{d_{i}}=\exp\left(-\frac{1}{\sqrt{\beta}}\sum_{k=i}^{2n}x_{k}^{-1/2}(\tilde{g}_{k}\sqrt{h})\right)$.
\item $\tilde{g}_{1},\dots,\tilde{g}_{2n}$ are independent, and, for any
$\varepsilon>0$, the random variables $\tilde{g}_{\lceil\varepsilon/h\rceil},\dots,\tilde{g}_{2n-1}$
have mean $O(\sqrt{h})$ and standard deviation $1+O(h)$, uniformly.
\end{enumerate}
\end{thm}
The point of (3) is that the sequence $\frac{1}{\sqrt{h}}\tilde{g}_{1},\dots,\frac{1}{\sqrt{h}}\tilde{g}_{2n}$
is a discretization of white noise. Hence, the expression for $e^{d_{i}}$
in (2) is a discretization of $\psi(x_{i})$.

\begin{proof}
Conclusions (1) and (2) are simply restatements of facts from (\ref{eq:identityformhard}).
For (3), the independence of $\tilde{g}_{1},\dots,\tilde{g}_{2n}$
was already established in the context of (\ref{eq:identityformhard}).
For the asymptotic mean and standard deviation, use the asymptotics
for chi-distributed random variables from the proof of the previous
theorem.
\end{proof}
Hence, $M_{\text{hard}}^{\beta,a}$ can be viewed as a finite difference
scheme for the stochastic Bessel operator,\[
M_{\text{hard}}^{\beta,a}\;\sim\; e^{D_{\text{even}}}M_{\text{hard}}^{\infty,a}e^{-D_{\text{odd}}}\;\stackrel{n\rightarrow\infty}{\rightsquigarrow}\;\psi\mathcal{J}_{a}^{\infty}\psi^{-1}\;\sim\;\mathcal{J}_{a}^{\beta}.\]

\subsubsection{Jacobi $\rightarrow$ Bessel}

\begin{defn}
The $n$-by-$n$ \emph{$\beta$-Jacobi matrix model scaled at the
hard edge} is\[
J_{\text{hard}}^{\beta,a,b}\sim\frac{1}{h}FB_{22}^{\beta,a,b}F,\]
in which $h=\frac{1}{2n+a+b+1}$ and $B_{22}^{\beta,a,b}$ is the
bottom-right block of the $2n$-by-$2n$ $\beta$-Jacobi matrix model
$J^{\beta,a,b}$. $F$ and $\Omega$ are defined in Section \ref{section:backgroundfinitediffs}.
\end{defn}
As $n\rightarrow\infty$, the small singular values of $J_{\text{hard}}^{\beta,a,b}$
display hard edge behavior. We explain this fact by interpreting the
rescaled matrix model as a finite difference scheme for the stochastic
Bessel operator in Liouville normal form with type (i) boundary conditions.
First, though, a lemma is required.

\begin{lem}
Suppose that $\theta$ is a random angle in $[0,\frac{\pi}{2}]$ whose
distribution is defined by $\cos^{2}\theta\sim\betapdf(c,d)$. Then
$\log(\tan^{2}\theta)$ has mean $\frac{\Gamma'(d)}{\Gamma(d)}-\frac{\Gamma'(c)}{\Gamma(c)}$
and variance $\frac{\Gamma(d)\Gamma''(d)-\Gamma'(d)^{2}}{\Gamma(d)^{2}}+\frac{\Gamma(c)\Gamma''(c)-\Gamma'(c)^{2}}{\Gamma(c)^{2}}$.
\end{lem}
\begin{proof}
$\tan^{2}\theta=\frac{1-\cos^{2}\theta}{\cos^{2}\theta}$ has a beta-prime
distribution with parameters $d$, $c$. Hence, $\tan^{2}\theta$
has the same distribution as a ratio of independent chi-square random
variables, with $2d$ degrees of freedom in the numerator and $2c$
degrees of freedom in the denominator. Let $X\sim\chi_{2d}^{2}$ and
$Y\sim\chi_{2c}^{2}$ be independent. Then the mean of $\log(\tan^{2}\theta)$
equals $E[\log X]-E[\log Y]=(\log2+\frac{\Gamma'(d)}{\Gamma(d)})-(\log2+\frac{\Gamma'(c)}{\Gamma(c)})=\frac{\Gamma'(d)}{\Gamma(d)}-\frac{\Gamma'(c)}{\Gamma(c)}$,
and the variance equals $\Var[\log X]+\Var[\log Y]=\frac{\Gamma(d)\Gamma''(d)-\Gamma'(d)^{2}}{\Gamma(d)^{2}}+\frac{\Gamma(c)\Gamma''(c)-\Gamma'(c)^{2}}{\Gamma(c)^{2}}$.
\end{proof}
\begin{thm}
\label{thm:betajacobitobessel}Let $J_{\text{hard}}^{\beta,a,b}$
be a matrix from the $n$-by-$n$ $\beta$-Jacobi matrix model scaled
at the hard edge. Adopting the notation of (\ref{eq:identityforjacobihard})
and Theorem \ref{thm:infjacobitobessel} and setting $\tilde{g}_{2i-1}=-\sqrt{(\beta x_{2i-1})/(2h)}\frac{1}{2}(\log(\tan^{2}\theta_{i})-\log(\tan^{2}\bar{\theta}_{i}))$
and $\tilde{g}_{2i}=\sqrt{(\beta x_{2i})/(2h)}\frac{1}{2}(\log(\tan^{2}\theta'_{i})-\log(\tan^{2}\bar{\theta}'_{i}))$,
we have
\begin{enumerate}
\item $J_{\text{hard}}^{\beta,a,b}\sim e^{D_{\text{even}}}J_{\text{hard}}^{\infty,a,b}e^{-D_{\text{odd}}}.$
\item $e^{d_{i}}=\exp\left(\left(-\sqrt{\frac{2}{\beta}}\sum_{k=i}^{2n-1}x_{k}^{-1/2}(\tilde{g}_{k}\sqrt{h})\right)+R\right)$,
in which $R=-(\log s_{n}-\log\bar{s}_{n})$ if $i=1$, $R=-(\log s'_{j-1}-\log\bar{s}'_{j-1})-(\log s_{n}-\log\bar{s}_{n})$
if $i=2j-1$ is odd and greater than one, or $R=(\log s_{j}-\log\bar{s}_{j})-(\log s_{n}-\log\bar{s}_{n})$
if $i=2j$ is even.
\item $\tilde{g}_{1},\dots,\tilde{g}_{2n-1}$ are independent, and, for
any $\varepsilon>0$, the random variables $\tilde{g}_{\lceil\varepsilon/h\rceil},\dots,\tilde{g}_{2n-1}$
have mean $O(\sqrt{h})$ and standard deviation $1+O(h)$, uniformly. 
\end{enumerate}
\end{thm}
The point of (3) is that the sequence $\frac{1}{\sqrt{h}}\tilde{g}_{1},\dots,\frac{1}{\sqrt{h}}\tilde{g}_{2n-1}$
is a discretization of white noise. Hence, the expression for $e^{d_{i}}$
in (2) is a discretization of $\psi(x_{i})^{\sqrt{2}}$. (The remainder
term $R$ has second moment $O(h)$ and is considered negligible compared
to the sum containing $2n-i$ terms of comparable magnitude.)

\begin{proof}
Conclusion (1) is direct from (\ref{eq:identityforjacobihard}). 

Now, we prove conclusion (2). According to (\ref{eq:identityforjacobihard}),
when $i=2j$ is even,\begin{multline*}
d_{2j}=-\sum_{k=j+1}^{n}(\log c_{k}-\log\bar{c}_{k})+\sum_{k=j}^{n-1}(\log s_{k}-\log\bar{s}_{k})\\
+\sum_{k=j}^{n-1}(\log c'_{k}-\log\bar{c}'_{k})-\sum_{k=j}^{n-1}(\log s'_{k}-\log\bar{s'}_{k}).\end{multline*}
Compare with\begin{multline*}
-\sqrt{\frac{2}{\beta}}\sum_{k=i}^{2n-1}x_{k}^{-1/2}(\tilde{g}_{k}\sqrt{h})=-\sum_{k=j+1}^{n}(\log c_{k}-\log\bar{c}_{k})+\sum_{k=j+1}^{n}(\log s_{k}-\log\bar{s}_{k})\\
+\sum_{k=j}^{n-1}(\log c'_{k}-\log\bar{c}'_{k})-\sum_{k=j}^{n-1}(\log s'_{k}-\log\bar{s}'_{k}).\end{multline*}
The remainder term $R$ is designed to cancel terms that occur in
one expression but not in the other. The argument for odd $i$ is
similar.

The asymptotics in conclusion (3) can be derived from the explicit
expressions in the previous lemma. The details are omitted.
\end{proof}

\subsubsection{Overview of finite difference schemes for the stochastic Bessel operator}

Considering Theorems \ref{thm:inflaguerreltobessel}, \ref{thm:inflaguerremtobessel},
\ref{thm:infjacobitobessel}, \ref{thm:betalaguerretobessel}, \ref{thm:betalaguerremtobessel},
and \ref{thm:betajacobitobessel},

\begin{enumerate}
\item $L_{\text{hard}}^{\beta,a}$ discretizes $\mathcal{J}_{a}^{\beta}$
with type (i) boundary conditions, for finite and infinite $\beta$.
\item $M_{\text{hard}}^{\beta,a}$ discretizes $\mathcal{J}_{a-1}^{\beta}$
with type (ii) boundary conditions, for finite and infinite $\beta$.
\item $J_{\text{hard}}^{\beta,a,b}$ discretizes $\tilde{\mathcal{J}}_{a}^{\beta}$
with type (i) boundary conditions, for finite and infinite $\beta$.
\end{enumerate}
Based on these observations and the fact that the small singular values
of $L_{\text{hard}}^{\beta,a}$, $M_{\text{hard}}^{\beta,a}$, and
$J_{\text{hard}}^{\beta,a,b}$ approach hard edge distributions as
$n\rightarrow\infty$, we pose the following conjecture.

\begin{conjecture}
Under type (i) boundary conditions, the $k$th least singular value
of the stochastic Bessel operator follows the $k$th hard edge distribution
with parameters $\beta$, $a$. Under type (ii) boundary conditions,
the hard edge distribution has parameters $\beta$, $a+1$. This is
true both for the original form, $\mathcal{J}_{a}^{\beta}$, and for
Liouville normal form, $\tilde{\mathcal{J}}_{a}^{\beta}$.
\end{conjecture}

\section{Numerical evidence}

\subsection{\label{section:airyrayleighritz}Rayleigh-Ritz method applied to
the stochastic Airy operator}

This section provides numerical support for the claim that stochastic
Airy eigenvalues display soft edge behavior. Up until now, our arguments
have been based on the method of finite differences. In this section,
we use the Rayleigh-Ritz method.

To apply Rayleigh-Ritz, first construct an orthonormal basis for the
space of $L^{2}((0,\infty))$ functions satisfying the boundary conditions
for $\mathcal{A}^{\beta}$. The obvious choice is the sequence of
eigenfunctions of $\mathcal{A}^{\infty}$. These functions are $v_{i}(x)=\frac{1}{\Ai'(\zeta_{i})}\Ai(-x+\zeta_{i})$,
$i=1,2,3,\dots,$ in which $\zeta_{i}$ is the $i$th zero of Airy's
function $\Ai$. Expanding a function $v$ in this basis, $v=c_{1}v_{1}+c_{2}v_{2}+\cdots$,
the quadratic form $\langle v,\mathcal{A}^{\beta}v\rangle$ becomes\begin{multline*}
\langle v,\mathcal{A}^{\beta}v\rangle=\sum_{i,j\geq1}\left(\langle c_{i}v_{i},\mathcal{A}^{\infty}c_{j}v_{j}\rangle+\frac{2}{\sqrt{\beta}}\int_{0}^{\infty}(c_{i}v_{i})(c_{j}v_{j})dB\right)\\
=\sum_{i,j\geq1}\left(-c_{i}^{2}\zeta_{i}\delta_{ij}+\frac{2}{\sqrt{\beta}}c_{i}c_{j}\int_{0}^{\infty}v_{i}v_{j}dB\right).\end{multline*}
 Note that the stochastic integral $\int_{0}^{\infty}v_{i}v_{j}dB$
is well defined, and its value does not depend on specifying an It\^{o}
or Stratonovich interpretation, because $v_{i}$ and $v_{j}$ are
well behaved and not random. (In fact, the joint distribution of the
stochastic integrals is a multivariate Gaussian, whose covariance
matrix can be expressed in terms of Riemann integrals involving Airy
eigenfunctions.) Introducing the countably infinite symmetric $K$,
\[
K=\left(-\zeta_{i}\delta_{ij}+\frac{2}{\sqrt{\beta}}\int_{0}^{\infty}v_{i}v_{j}dB\right)_{i,j=1,2,3,\dots},\]
the quadratic form becomes $c^{T}Kc$, in which $c=(c_{1},c_{2},c_{3},\dots)^{T}$,
the vector of coefficients in the basis expansion.

According to the variational principle, the least eigenvalue of $\mathcal{A}^{\beta}$
equals $\inf_{\| v\|=1}\langle v,\mathcal{A}^{\beta}v\rangle$, which
equals $\min_{\| c\|=1}c^{T}Kc$, which equals the minimum eigenvalue
of $K$. This suggests a numerical procedure. Truncate $K$, taking
the top-left $l$-by-$l$ principal submatrix, and evaluate the entries
numerically. Then compute the least eigenvalue of this truncated matrix.
This is the Rayleigh-Ritz method.

The histograms in Figure \ref{fig:airyrayleighritz} were produced
by running this procedure over $10^{5}$ random samples, discretizing
the interval $(0,86.9)$ with a uniform mesh of size 0.05 and truncating
$K$ after the first 150 rows and columns. The histograms match the
soft edge densities well, supporting the claim that the least eigenvalue
of $\mathcal{A}^{\beta}$ exhibits soft edge behavior.

\subsection{\label{section:besselrayleighritz}Rayleigh-Ritz method applied to
the stochastic Bessel operator}

Now consider applying the Rayleigh-Ritz method to the stochasic Bessel
operator in Liouville normal form with type (i) boundary conditions.
Liouville form is well suited to numerical computation because the
singular functions are well behaved near the origin for all $a$.
We omit consideration of type (ii) boundary conditions for brevity.

Two orthonormal bases play important roles, one consisting of right
singular functions and the other consisting of left singular functions
of $\tilde{\mathcal{J}}_{a}^{\infty}$, from (\ref{eq:besselliouvillesvd}).
For $i=1,2,3,\dots$, let $v_{i}(x)$ be the function $\sqrt{x}j_{a}(\xi_{i}x)$,
normalized to unit length, and let $u_{i}(x)$ be the function $\sqrt{x}j_{a+1}(\xi_{i}x)$,
normalized to unit length, in which $\xi_{i}$ is the $i$th zero
of $j_{a}$.

The smallest singular value of $\tilde{\mathcal{J}}_{a}^{\beta}$
is the minimum value for $\frac{\|\tilde{\mathcal{J}}_{a}^{\beta}v\|}{\| v\|}=\frac{\|\psi^{\sqrt{2}}\tilde{\mathcal{J}}_{a}^{\infty}\psi^{-\sqrt{2}}v\|}{\| v\|}$.
Expressing $v$ as $v=f\psi^{\sqrt{2}}$ and expanding $f$ in the
basis $v_{1},v_{2},v_{3},\dots$ as $f=\sum_{i=1}^{\infty}c_{i}v_{i}$,
the norm of $\|\tilde{\mathcal{J}}_{a}^{\beta}v\|$ becomes\begin{multline*}
\|\tilde{\mathcal{J}}_{a}^{\beta}v\|=\|\psi^{\sqrt{2}}\tilde{\mathcal{J}}_{a}^{\infty}f\|\\
=\left(\int_{0}^{1}\psi^{2\sqrt{2}}\left(\sum_{i=1}^{\infty}\xi_{i}c_{i}u_{i}\right)^{2}dt\right)^{1/2}=\left(\sum_{i,j\geq1}c_{i}c_{j}\xi_{i}\xi_{j}\int_{0}^{1}\psi^{2\sqrt{2}}u_{i}u_{j}dt\right)^{1/2}.\end{multline*}
In terms of the countably infinite symmetric matrix $K$,\[
K=\left(\xi_{i}\xi_{j}\int_{0}^{1}\psi^{2\sqrt{2}}u_{i}u_{j}dt\right)_{i,j\geq1},\]
we have $\|\tilde{\mathcal{J}}_{a}^{\beta}v\|=(c^{T}Kc)^{1/2}$, in
which $c=(c_{1},c_{2},c_{3},\dots)^{T}$. For the norm of $v$, we
find\begin{multline*}
\| v\|=\| f\psi^{\sqrt{2}}\|=\left(\int_{0}^{1}\psi^{2\sqrt{2}}\left(\sum_{i=1}^{\infty}c_{i}v_{i}\right)^{2}dt\right)^{1/2}\\
=\left(\sum_{i,j\geq1}c_{i}c_{j}\int_{0}^{1}\psi^{2\sqrt{2}}v_{i}v_{j}dt\right)^{1/2}=(c^{T}Mc)^{1/2},\end{multline*}
in which $c$ is defined as above and $M$ is the countably infinite
symmetric matrix\[
M=\left(\int_{0}^{1}\psi^{2\sqrt{2}}v_{i}v_{j}dt\right)_{i,j\geq1}.\]

The least singular value of $\tilde{\mathcal{J}}_{a}^{\beta}$ equals
$\min\frac{\|\tilde{\mathcal{J}}_{a}^{\beta}v\|}{\| v\|}$, which
equals $\min\left(\frac{c^{T}Kc}{c^{T}Mc}\right)^{1/2}$, which equals
the square root of the minimum solution $\lambda$ to the generalized
eigenvalue problem $Kc=\lambda Mc$. To turn this into a numerical
method, simply truncate the matrices $K$ and $M$, and solve the
resulting generalized eigenvalue problem.

The histograms in Figure \ref{fig:besselrayleighritz} were produced
using this method on $10^{4}$ random samples of the stochastic Bessel
operator, discretizing the interval $(0,1)$ with a uniform mesh of
size $0.001$ and truncating the matrices $K$ and $M$ after the
first 75 rows and columns. The histograms match the hard edge densities
well, supporting the claim that the least singular value of the stochastic
Bessel operator follows a hard edge distribution.

\subsection{\label{sec:smoothness}Smoothness of eigenfunctions and singular
functions}

Up to this point, we have proceeded from random matrices to stochastic
operators. In this section, we reverse direction, using stochastic
operators to reveal new facts about random matrices. Specifically,
we make predictions regarding the {}``smoothness'' of Hermite eigenvectors
and Jacobi CS vectors, using the stochastic operator approach. Verifying
the predictions numerically provides further evidence for the connection
between classical random matrix models and the stochastic Airy and
Bessel operators.

First, consider the eigenfunctions of the stochastic Airy operator.
The $k$th eigenfunction is of the form $f_{k}\phi$, in which $f_{k}\in C^{2}((0,\infty))$
and $\phi\in C^{3/2-}((0,\infty))$ is defined by (\ref{eq:phi}).
In light of the claim that $H_{\text{soft}}^{\beta}$ encodes a finite
difference scheme for $\mathcal{A}^{\beta}$, the $k$th eigenvector
of $H_{\text{soft}}^{\beta}$ should show structure indicative of
the $k$th eigenfunction $f_{k}\phi$ of $\mathcal{A}^{\beta}$. For
a quick check, consider the ratio of two eigenfunctions/eigenvectors.
The $k$th eigenfunction of $\mathcal{A}^{\beta}$ is of the form
$f_{k}\phi$, which does not have a second derivative (with probability
one) because of the irregularity of Brownian motion. However, the
ratio of the $k$th and $l$th eigenfunctions is $\frac{f_{k}\phi}{f_{l}\phi}=\frac{f_{k}}{f_{l}}$,
which, modulo poles, has a continuous second derivative. Therefore,
we expect the entrywise ratio between two eigenvectors of $H_{\text{soft}}^{\beta}$,
$L_{\text{soft}}^{\beta,a}$, or $M_{\text{soft}}^{\beta,a}$ to be
{}``smoother'' than a single eigenvector. Compare Figure \ref{fig:predictionforhermite}.

Next, consider the singular functions of the stochastic Bessel operator
in Liouville normal form. The $k$th right singular function is of
the form $f_{k}\psi^{\sqrt{2}}$ and the $k$th left singular function
is of the form $g_{k}\psi^{-\sqrt{2}}$, in which $f_{k},g_{k}\in C^{1}((0,1))$
and $\psi\in C^{1/2-}((0,1))$ is defined by (\ref{eq:psi}). The
situation is similar to the Airy case. Any one singular function may
not be differentiated in the classical sense, because of the irregularity
of Brownian motion. However, the ratio of two singular functions is
smooth. We expect the singular vectors of $L_{\text{hard}}^{\beta,a}$,
$M_{\text{hard}}^{\beta,a}$, and $J_{\text{hard}}^{\beta,a,b}$ to
show similar behavior. Compare Figure \ref{fig:predictionforjacobi}.

\section{Preview of the stochastic sine operator}

We have seen that the eigenvalues of the stochastic Airy operator
display soft edge behavior, and the singular values of the stochastic
Bessel operator display hard edge behavior. Is there a stochastic
differential operator whose eigenvalues display bulk behavior? Because
of the role of the sine kernel in the bulk spacing distributions,
it may be natural to look for a \emph{stochastic sine operator}. In
fact, \cite{mythesis} provides evidence that an operator of the form\begin{equation}
\left[\begin{array}{c|c}
 & -\frac{d}{dx}\\
\hline \frac{d}{dx}\end{array}\right]+\left[\begin{array}{c|c}
\text{``noise''} & \text{``noise''}\\
\hline \text{``noise''} & \text{``noise''}\end{array}\right]\label{eq:stochasticsine}\end{equation}
may be the desired stochastic sine operator. This operator is discovered
by scaling the Jacobi matrix model at the center of its spectrum,
and an equivalent operator, up to a change of variables, is discovered
by scaling the Hermite matrix model at the center of its spectrum.
The exact nature of the noise terms in (\ref{eq:stochasticsine})
is not completely understood at this point. A change of variables
analogous to those that transform (\ref{eq:intuitiveairy}) to (\ref{eq:technicalairy})
and (\ref{eq:intuitivebessel}) to (\ref{eq:technicalbessel}) would
be desirable.

\begin{acknowledgement*}
The authors thank Ioana Dumitriu for the inspiration provided by her
ideas. Figure \ref{fig:softhardbulk} was produced with the help of
Per-Olof Persson's software.
\end{acknowledgement*}
\bibliographystyle{plain}
\bibliography{thesis}

\end{document}